\documentclass[preprint,authoryear,12pt,review]{elsarticle}

\usepackage{graphicx}
\usepackage{amssymb}
\usepackage{amsmath}
\pdfoutput=1

\journal{Journal of Neuroscience Methods}

\usepackage{color} %inclui suporte a cores no texto para comando \needref
\usepackage{url}
\usepackage[latin1]{inputenc}
\usepackage[T1]{fontenc}

\def\virg{\ \textnormal{,}}
\def\ponto{\ \textnormal{.}}
\newcommand{\func}[1]{\textnormal{#1}}
\def\lmean{\left\langle}
\def\rmean{\right\rangle}
\def\atanh{\textnormal{atanh}}

\begin{document}

\begin{frontmatter}



\title{A Brief History of Excitable Map-Based Neurons and Neural Networks}


\author[endUFSC]{M. Girardi-Schappo}
\author[endUFSC]{M. H. R. Tragtenberg}
\author[endUSP]{O. Kinouchi}
\address[endUFSC]{Departamento de F\'isica, Universidade Federal de Santa Catarina, 88040-900, Florian\'opolis, Santa Catarina, Brazil}
\address[endUSP]{Departamento de F\'isica, FFCLRP, Universidade de S\~ao Paulo, 14040-900, Ribeir\~ao Preto, S\~ao Paulo, Brazil}
%\email{marcelotragtenberg@gmail.com}

\begin{abstract}
This review gives a short historical account of the excitable maps approach for modeling neurons and neuronal networks. Some early models, due to 
\citet{pasemann1993}, \citet{chialvo1995} and \citet{modeloKT}, are compared with more recent proposals by \citet{rulkovMapa} and \citet{izhikevichModel}.
We also review map-based schemes for electrical and chemical synapses and some recent findings as critical avalanches in map-based neural networks.
We conclude with suggestions for further work in this area like more efficient maps, compartmental modeling and close dynamical comparison with
conductance-based models.    
\end{abstract}


\end{frontmatter}

% \linenumbers

\section{Introduction}

\label{sec:intro}

The number of neurons in the human brain (86 billions \citep{suzanaBrain}) 
is near six times the number of trees in Amazonia. 
So, brain modelers must not forget that their job is comparable to modeling patches of
Amazonia, a staggering task. Since well developed models for single neurons already exist \citep{genesisBook,neuronBook}, 
with complex dendritic geometry and tens
of equations and parameters \citep{abbottNeuro}, 
it is not obvious what modeling level we should use in general.

As the proverbial forest not seen because of the trees,
the detailed study of singular neurons is an interesting subject \textit{per se} but perhaps
not necessary to understand the macroscopic dynamics and function of neuronal networks.
Indeed, neuronal networks present collective phenomena,
like synchronization, waves and avalanches, with regimes separated by global bifurcations or
\textit{phase transitions}, that cannot be studied in small neuronal populations. 
The history of neuronal networks modeling is marked by this trade-off between 
analytical/computational tractability and biological realism.

Since the connection between neurons is only sensitive to
the action potentials that arrive at (electrical or chemical) synapses, the important thing
is to model the dynamics of these action potentials (their frequency or inter-spike intervals, if they
come in bursts or single events etc.). The emphasis in modeling the transmembrane voltage dynamical
behavior is called a phenomenological approach (in contrast to a mechanistic or biophysical approach),
leads to a class of neuron models where map-based neurons are a new and promising tool. 
This paper gives a brief account of the pionnering proposals of neuronal maps due to
\citet{chialvo1995}, \citet{pasemann1993,pasemann1997}, \citet{modeloKT} and \citet{modeloKTz2001} and compare
them with more recent proposals due to \citet{rulkovChaotic,rulkovMapa} and \citet{izhikevichModel,izhikevichMapas}.
%\cite{Ibarz2011, Izhikevich2011}.
  
There is two main routes to achieve map-based neuron models with realistic dynamical properties.
The first one is to start from Hodgkin-Huxley (HH) type models, composed by coupled 
nonlinear ordinary differential equations (ODE) which are
already a simplification (due to spatial discretization) of full partial differential equations
that describes the neuron membrane. Computational neuroscience models 
based on the HH formalism, called conductance-based neurons, is a well developed subject \citep{abbottNeuro}, 
but suffers from some limitations \citep{deSchutterBook}:
\begin{itemize}
\item The HH-type models consist in several nonlinear coupled EDOs: the simulation of a single
neuron is orders of magnitude more costly than simplified neuron models;
\item The biophysical data for constraining the parameter values (like capacitances, axial resistances,
density of ion channels, etc.) is scarce and often obtained from diverse preparations (different animals,
in vitro experiments etc.). Most of the parameter ranges used in simulations are simply informed guesses.
\item The remaining parameter space of these models is huge and suffer from the so called curse of dimensionality
\citep{curseDim}. It is very costly to trace full phase diagrams, since with $P$ parameters, for example, 
we can have $P(P-1)$ parameter planes. The minimal model of Hodgkin-Huxley, with only two 
active ion channels, has at least $P = 40$ parameters \citep{abbottNeuro}.
\item The set of parameters to be used for reproducing a given firing pattern is subdetermined. 
This means that the same dynamical behavior can be achieved by different sets of parameters. 
Adjusting these parameters to the known neuron behavior is susceptible of overfitting: the model
reproduces the given data but do not generalizes well, for example, for different input situations.
\end{itemize} 

In order to deal with these drawbacks, we may opt to reproduce the dynamical behavior of neurons 
instead of reproducing the involved biophysical mechanisms (mechanistic modeling).
Starting from a complicated HH-model, perhaps even a multicompartimental model,
we can perform a sequence of simplifications more or less justified
in order to obtain simpler models with fewer equations and lumped parameters \citep{deSchutterBook}.
Examples of these reduced ODE's based models are the FitzHugh-Nagumo excitable neuron \citep{fitzhugh,nagumo},
the Hindmarsh-Rose bursting neuron \citep{modeloHR} and the Izhikevitch model \citep{izhikevichModel}.
If we numerically integrate these ODEs with the Euler method with a large time step, we can arrive to
maps with similar dynamical properties as the original systems \citep{rulkovMapa,izhikevichWhich}.

Phenomenological modeling can start the other way around. This occurs because 
the phenomena to be studied sets the level of modeling. Continuing with our forest modeling analogy,
if our interest is to study a single tree (or neuron), a biophysical HH-like modeling is desirable. 
But if we want to understand, say, the propagation
of a forest fire, the modeling of the tree biophysics is mostly immaterial, and trees could be 
represented by sites with two states ($0 =$ normal, $1 =$ burnt) \citep{christensenFFM}. In the same vein, \citet{McCulloch1943}
proposed a binary threshold neuron, whose state is given by $0 =$ rest and $1 =$ firing. 
With this method, one starts with discrete time systems and searches for increasing complexity 
until achieve dynamical models that reproduce the full phenomenology of neuronal dynamics. 

Both approachs tend to converge at a middle ground formalism: dynamical systems
with discrete space and discrete time, but with continuous state variables, that is, dynamical maps \citep{ibarzMapas}.
Neuronal networks composed by these maps will be an instance of coupled maps lattices (CMLs) \citep{kanekoCMLbook,kanekoCML}.
In this paper, we review the early proposals of map-based neuron models and the coupling schemes
used to create such coupled maps networks. We also suggest some unexplored research topics
that could be examined both with conductance-based neurons and map-based  neurons,
in order to stimulate computational neuroscientists to use both approachs in a synergetic way.

This review is intended to organize the map-based neuron models in sequential time (Sec. \ref{sec:hist}),
highlighting two families of map-based modeling: (I) back from \citet{McCulloch1943} approach to the \citet{modeloKT}
and its extension \citep{modeloKTz2001} and (II) back from \citet{chialvo1995} to \citet{rulkovMapa} and \citet{izhikevichModel}.
Then, we perform a short computational comparison of the main neuronal models (Sec. \ref{sec:discus}).
The main purpose of Sec. \ref{sec:coup} is to neatly list the most relevant couplings which may be used to link
maps into networks, only pointing to the most prominent results obtained with them. We finally terminate
the review with some important remarks in Sec. \ref{sec:conc}.

\section{History of Map-based Neurons}
\label{sec:hist}

This section is devoted to draw a line which connects the early modeling of neurons, as state machines, to the most recent and powerful models, which are
dynamical systems on their own, presenting their most relevant features and reviewing some models that are still not well known, although they have been used
recently to model neural networks.

The generalized mathematical form of any map is:
\begin{equation}
\label{eq:generalMap}
\vec{x}(t+1) = \vec{F}\left[\vec{x}(t)\right]\virg
\end{equation}
where $\vec{F}:\mathbb{R}^n\rightarrow\mathbb{R}^n$ is any vector function and we are assuming that the curve given by the set of values $\left\{\vec{x}(t)\right\}$
defines the temporal evolution of some quantity. In the case of neurons,
each component of
the vector $\vec{x}(t)$ accounts for a relevant neuronal quantity.

Generally, the first component, $x^1(t)$, is the membrane potential (the fast variable) and
the second component, $x^2(t)$, may be the slow inward and outward currents or an auxiliary variable. When present, the third component, $x^3(t)$,
accounts for the slow currents. For the sake of simplicity, we define $x(t)\equiv x^1(t)$, $y(t)\equiv x^2(t)$ and $z(t)\equiv x^3(t)$ throughout this paper.
The subscript index is intended to identify elements of a network.

\subsection{Early History}
\label{sec:early}

\citet{McCulloch1943} formal neuron can be viewed, if coupled to $N$ presynaptic neurons with parallel update, as a discrete time
dynamical system \citep{little1974}:
\begin{equation}
\label{eq:MPModel}
x_i(t+1) = H\left(\sum_{j\neq i}^N J_{ij} x_j(t) - \theta + I_i(t) \right)
\end{equation}
where the Heaviside (step) function gives $H(y) = 1$ if $y\ge 0$ (zero otherwise), $I_i(t)$ is the external input and 
$\theta$ is a firing threshold.

In the statistical physics community, due to symmetry motivations, it is common to scale the neuron
output as a sign function with values $\pm 1$, that is, $S(y) = 2 H(y) -1$.

The isolated McCulloch-Pitts neuron has no intrinsic dynamics (notice that the input sum is over $j\neq i$).
However we can introduce a dependence on the past values of its variable, as in the \citet{caianello1961} equations:
\begin{equation}
\label{eq:caianelloModel}
x_i(t+1) = H\left[\sum_{n=0}^\tau\left( W_i^{(n)} x_i(t-n) + \sum_{j\neq i}^N J_{ij}^{(n)} x_j(t)\right) - \theta +I_i(t) \right], 
\end{equation}
where now the $W_i^{(n)}$ parameters weight the contributions of the delayed $x_i(t-n)$ values within a memory window $\tau$.
So, the isolated neuron ($J_{ij} = 0$) can present interesting dynamical behavior.
 
Indeed, \citet{nagumo1972} studied an isolated formal neuron with an exponential decay of refractory influence, that is,
\begin{equation}
\label{eq:NSModel}
x(t+1) = H\left(\sum_{n=0}^\tau W^{(n)} x(t-n)  - \theta +I(t) \right), 
\end{equation}
where $W^{(n)} = - \alpha w^n$,
with a decay constant $w$. It has been shown that almost all solutions of the Nagumo-Sato model are periodic 
and form a complete devil staircase -- chaotic solutions lie at the complementary zero measure cantor set
\citep{aihara2010}.
 
\begin{figure}[b!]
	\begin{center}
	\includegraphics[width=80mm]{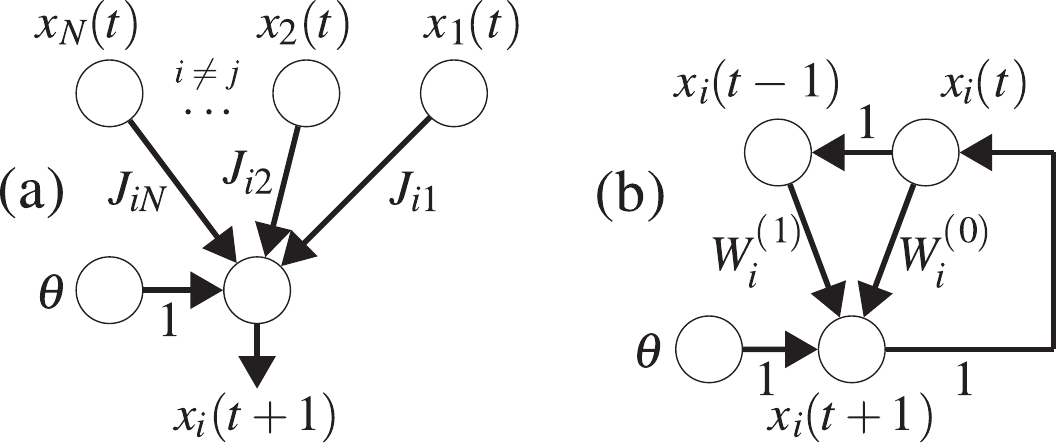}
	\end{center}
	\caption{\label{fg:perceptron}(a) Scheme of a single-layer perceptron (Eq. \ref{eq:MPModel}) and (b) the so-called dynamical perceptron (Eq. \ref{eq:dynPerceptron} with $\tau=1$). In both cases, the output $x_i(t+1)$ may be calculated by a continuous function or by a discret step function.}
\end{figure}

Since the $x(t)$ variable has a discrete set of values, this kind of formal neurons corresponds to cellular automata, not to continuous maps.
But starting from the decade of 1980, modelers substituted  the discontinuous step or sign functions by continuous ones:
\begin{equation}
\label{eq:dynPerceptron}
x(t+1) = F\left(\sum_{n=0}^\tau W^{(n)} x(t-n)  - \theta +I(t) \right), 
\end{equation}
where, for example, the transfer function is a logistic function $F(y) = [1+\exp(-\gamma y)]^{-1} $ or a hyperbolic tangent
$F(y) = \tanh(y/T)$ \citep{Hopfield1984,Aihara1990}, where $\gamma$ and $1/T$ are gain parameters. 
Now, the $x(t+1) = F\left(x(t), x(t-1), ..., x(t-\tau) \right)$ is a $\tau + 1$ dimensional map.
Fig. \ref{fg:perceptron} illustrates the differences between Eqs. \ref{eq:MPModel} and \ref{eq:dynPerceptron} for $\tau=1$.

The case $\tau =0$ with a logistic function $F(y)$, that is, $x(t+1) = F\left(\gamma (x(t) + I(t) -\theta) \right)$ 
was examined by \citet{pasemann1993}. In the plane ($\gamma$, $H=I-\theta$) it presents only
three behaviors: a fixed point phase, a period two phase and a bistability (two fixed points) phase \citep{pasemann1993,modeloKT}.
A chaotic version of this neuron map has also been studied by \citet{pasemann1997}. 

Due to a lucky accident, the case $\tau = 1$ implements a powerful excitable element with a very rich behavior.
We call this system a second order dynamical perceptron or the \citet{modeloKT} map (also known in the statistical mechanics
literature as the YOS map \citep{yokoi1985,tragtenbergYokoi}). This excitable neuron model is reviewed in the next section.

So far all the models follow the same principle: they are built directly from discrete time dynamics.
However, neuronal models can be built the other way around: starting from continuous time differential equations,
like the HH model, and then simplify them to keep only their wanted features.
A result of this kind of simplification is proposed by \citet{chialvo1995} to study excitable systems (and, in particular, neural excitability):
\begin{equation}
\label{eq:chialvoModel}
\left.\begin{array}{l}
x(t+1) = [x(t)]^2\exp[y(t)-x(t)]+I(t)\\
y(t+1) = ay(t)-bx(t)+c
\end{array}\right.\virg
\end{equation}
where $a$, $b$ and $c$ are parameters and $I(t)$ may account for a bias membrane potential or external input. Chialvo inspired more
recent works, like the \citet{rulkovMapa} and the \citet{izhikevichModel} models, both studied in Sec. \ref{sec:recent}.

\subsection{KT and KTz Maps}
\label{sec:ktz}

The case $\tau=1$ with $\func{F}(y)=\tanh(y)$ was extensively studied in the context of Statistical Mechanics and resulted in many phase diagrams \citep{tragtenbergYokoi}. In order to build on these results, \citet{modeloKT}
imposed, in Eq. \ref{eq:dynPerceptron}, the change of parameters
$K\equiv -W^{(1)}/W^{(0)}$, $T\equiv 1/W^{(0)}$ and $H\equiv (\theta + W^{(0)} + W^{(1)})/W^{(0)}$. Rewriting Eq. \ref{eq:dynPerceptron}
leads us to the KT model:
\begin{equation}
\label{eq:KTModel}
\left.\begin{array}{l}
x(t+1) = \tanh\left(\dfrac{x(t) - Ky(t)+ H + I(t)}{T}\right)\virg\\
y(t+1) = x(t)\virg
\end{array}\right.
\end{equation}
where $-1<x(t)<+1$ represents the actual membrane potential of the neuron at time $t$ -- measured in \textit{timesteps} (ts). The term $I(t)$ corresponds to an external input. In section
\ref{sec:coup}, the coupling is done via $I(t)$ as well. Both $x(t)$, $I(t)$ and $t$ may be conveniently rescaled to any unit system, like
mili-Volts or
mili-seconds (ms). The authors assumed $1$ ts = $0.1$ ms whilst the membrane potential may be rescaled to fit a Hodgkin-Huxley
model, for instance, by relating the value of the membrane potential at the fixed point and at the peak of the spike in
both models.

\begin{figure}[h!]
	\begin{center}
	\includegraphics[width=50mm]{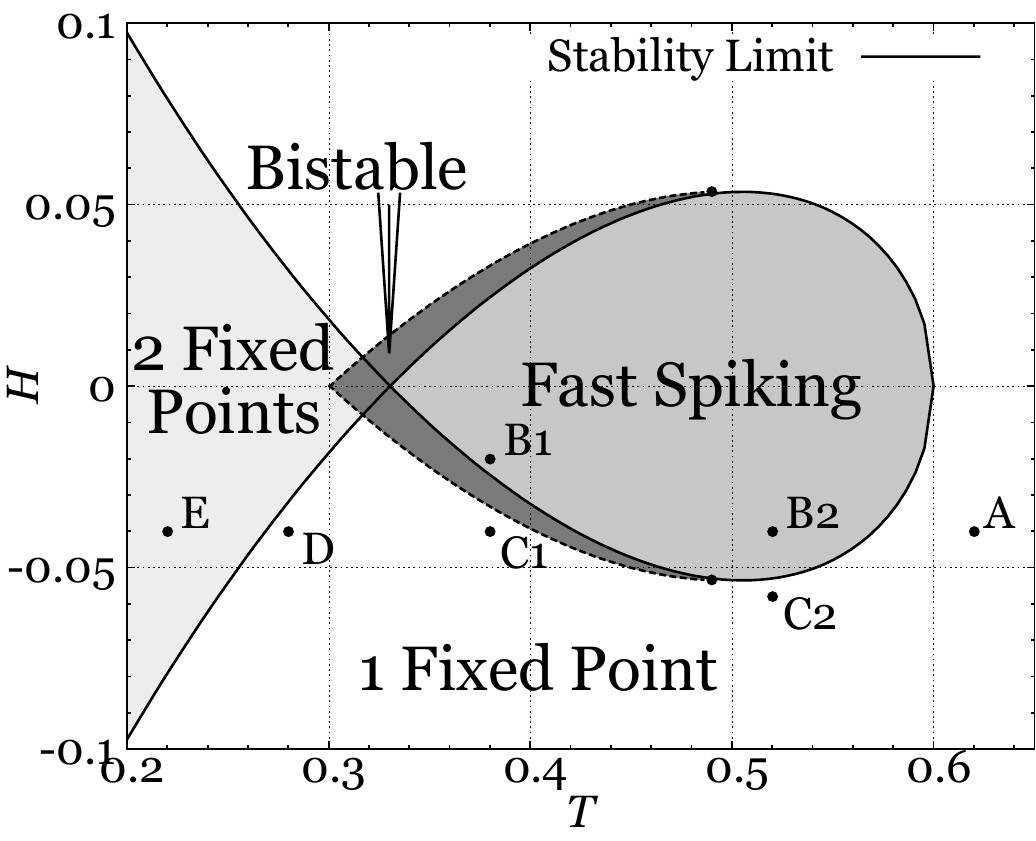}
	\end{center}
	\caption{\label{fg:KTPhaseDiag}KT model phase diagram for $K=0.6$. It allowed authors to list five different neuronal behaviors, labeled from A to E, discussed in the text. The Bistable region corresponds to coexistence of oscillatory with fixed point behaviors.}
\end{figure}

The authors reinterpreted the statistical mechanical phase diagrams, as shown in Fig. \ref{fg:KTPhaseDiag}, proposing that the relevant
neuronal behaviors are in the parameter subspace given by $0.5\leq K\leq1$, with $T$ and $H$ near the phase transitions from fixed point to
oscillatory behavior (i.e. fast spiking).

The stability limit of the model, the curve $H_s(K,T)$, separating the oscillating and the fixed point phases may be calculated
via a simple linear stability analysis, which yields:
\begin{equation}
\label{eq:KTStabLim}
H_s^\pm(K,T) = T\atanh(x^*_s) - (1-K)x^*_s\virg
\end{equation}
where $x^*_s$ is the fixed point over the line $H_s(K,T)$, given by:
\begin{equation}
\label{eq:KTStabFP}
x^*_s = \left\{
\begin{array}{ll}
\pm\sqrt{1-\dfrac{T}{1-K}} & \textnormal{, if }0<K\leq0.5\\
\pm\sqrt{1-\dfrac{T}{K}}   & \textnormal{, if }0.5<K\leq1
\end{array}
\right.\ponto
\end{equation}
Differently from the statistical mechanical model, the KT model may be extended for $T<0$ and $K<0$. It would result in a slightly
different $x^*_s$, although the analysis would keep the same \citep{modeloKT}. Thus, we will focus on positive parameters.

Notice that the parameter $H$ may be redefined as $\tilde{H}=H+I(t)$, so the effect of an external
input is to drag the solution $x(t)$ for Eq. \ref{eq:KTModel} along any vertical line (with fixed $T$) on Fig. \ref{fg:KTPhaseDiag}.
This is the basic mechanism of the KT model excitability, in which $I(t)$ is such that $H+I(t)>H_s(K,T)$ -- assuming
the neuron is adjusted below de bubble of Fig. \ref{fg:KTPhaseDiag}.
This allows the authors to list five distinct neuronal behaviors (labeled
from A to E in Fig. \ref{fg:KTPhaseDiag}). These behaviors are present in Hodgkin-Huxley-like neurons and in experimental setups \citep{modeloKT}:
\begin{itemize}
	\item Neuron A: Silent neuron -- it will not fire any action potential, regardless the external stimulus intensity -- no
	bifurcation occurs;
	\item Neuron B (1 and 2): Pacemaker neuron -- it is an autonomous oscillator, although external stimulation may bring the neuron to
	outside of the bulb in Fig. \ref{fg:KTPhaseDiag} temporarily via a Subcritical Neimarck-Sacker bifurcation (B1) or a Supercritical
	Neimarck-Sacker bifurcation (B2);
	\item Neuron C (1 and 2): Excitable neuron -- it will fire only for a stimulus greater than some threshold; the stimulus
	takes the neuron inside the bulb in Fig. \ref{fg:KTPhaseDiag} through a Subcritical Neimarck-Sacker Bifurcation (C1) or a
	Supercritical Neimarck-Sacker Bifurcation (C2). If a neuron C1 is adjusted over the Bistable region, then external stimuli may
	switch between fast spiking and fixed point response (phenomenon known as pacemaker activity annihilation);
	\item Neuron D: presents a single fixed point; however, a strong enough external current, $I(t)$, may transform it into a bistable
	neuron with two stable fixed points;
	\item Neuron E: presents two stable fixed points even without external input; the neuron may switch between states through some stimulation
	$I(t)$.
\end{itemize}

\begin{figure}[t!]
	\begin{center}
	\includegraphics[width=100mm]{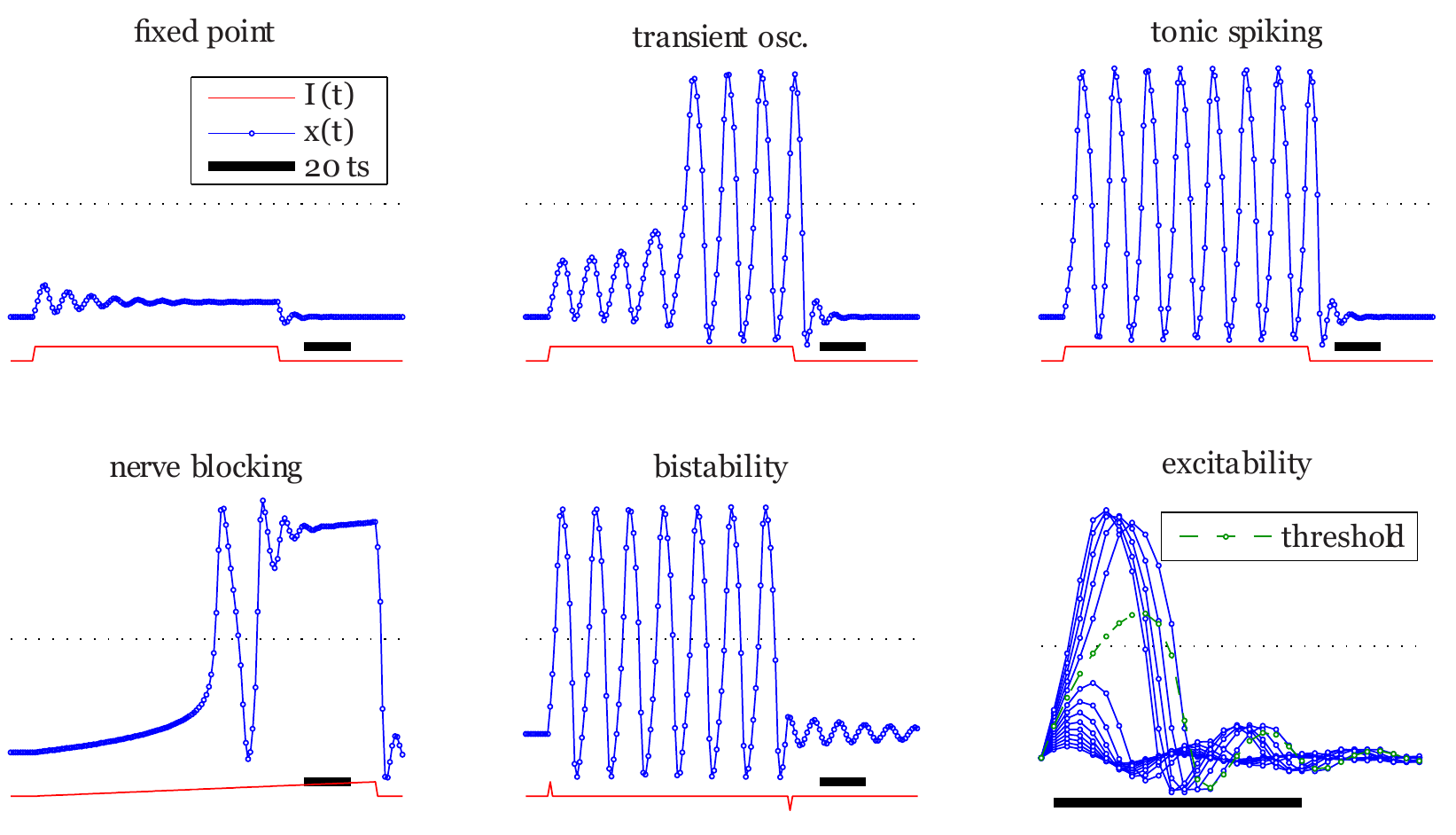}
	\end{center}
	\caption{\label{fg:KTTypeC}Excitable response of KT type C1 model. The solution $x(t)$ (--$\circ$--) corresponds only to the circle points (the lines are just guides so one can follow the sequence in which the map evolves through time). Solid line is external input and the bolder line segment corresponds to $20$ ts.}
\end{figure}

Type C1 neuron, specially, presents a rich repertoire of excitable responses (see Fig. \ref{fg:KTTypeC}). It also gave the authors 
insights about the mathematical nature of transient oscillations -- not explained, but experimentally observed by \citet{morrisLecar}: all behaviors
of type C1 are achieved because the neuron lies near a bifurcation point.

\begin{figure}[t!]
	\begin{center}
	\includegraphics[width=100mm]{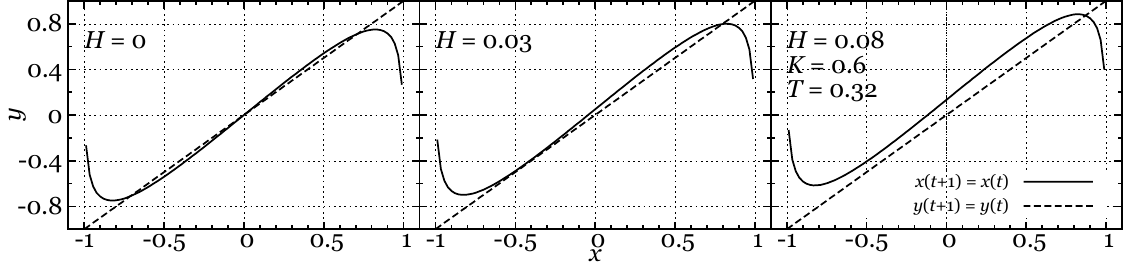}
	\end{center}
	\caption{\label{fg:KTNullclines}Effect of varying parameter $H$ on the nullclines of the KT model. Notice the similarity with the nullclines of the FitzHugh-Nagumo model.}
\end{figure}

KT model also presents a well defined excitability, as shown in the last panel of Fig. \ref{fg:KTTypeC} -- in which the neuron receives a delta current input $I(t) = I_0\delta_{t,t_0}$ in $t_0=0$ with increasing intensity $I_0$. The circle-dash line shows the approximate threshold $I_s$: any stronger stimulus causes a spike. The nullclines of the model are of the same shape as those of the FitzHugh-Nagumo model (see Fig. \ref{fg:KTNullclines}).

\begin{figure}[h!]
	\begin{center}
	\includegraphics[width=100mm]{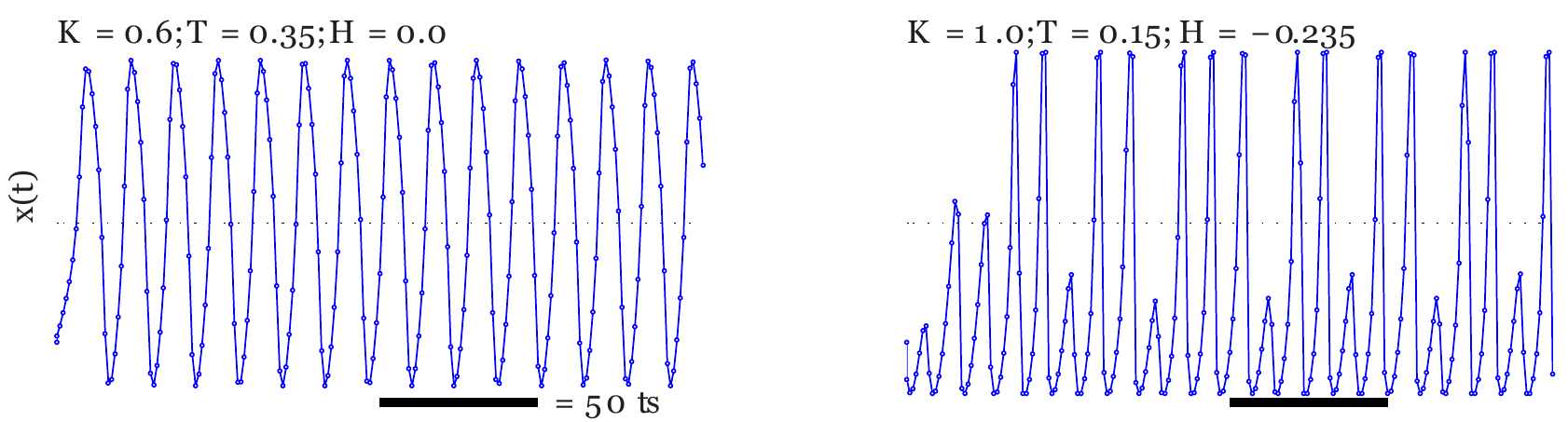}
	\end{center}
	\caption{\label{fg:KTFastSChaoS}Left: Fast spiking behavior (inside the bulb in Fig. \ref{fg:KTPhaseDiag}). Right: Chaotic spiking.}
\end{figure}

Besides excitability, KT model also presents regular and chaotic autonomous behavior (Fig. \ref{fg:KTFastSChaoS}). The authors also
proposed a mechanism to generate bursts based on the Hindmarsh-Rose model \citep{modeloKT}: by letting the parameter $H\equiv z(t)$, in Eq. \ref{eq:KTModel}, oscillate
slowly in time, with
\begin{equation}
\label{eq:KTeqForz}
z(t+1)=(1-\delta)z(t)-\lambda(x(t)-y(t))^2\virg
\end{equation}
where $\delta$ and $\lambda$ are paremeters that control the inward and the outward ionic currents, respectively. This modification allows the neuron to go back and forth
inside the bulb in Fig. \ref{fg:KTPhaseDiag}. This mechanism generates bursts when the transition induced
by the $z(t)$ dynamics is via a Subcritical Hopf Bifurcation (i.e. through the bistable region in Fig. \ref{fg:KTPhaseDiag}).

Furthermore, \citet{modeloKTRessonancia} showed that KT model, Eq. \ref{eq:KTModel}, may present \textit{stochastic resonance} (which is the occurrence of spiking
due to subthreshold stochastic stimulation). Beyond the usual stochastic resonance, the authors achieved autonomous and aperiodic
stochastic resonance as well.%As a matter of fact, the inside of the bulb in Fig. \ref{fg:KTPhaseDiag} is filled with many
%different oscillating phases; each with a given period.

\begin{figure}[h!]
	\begin{center}
	\includegraphics[width=60mm]{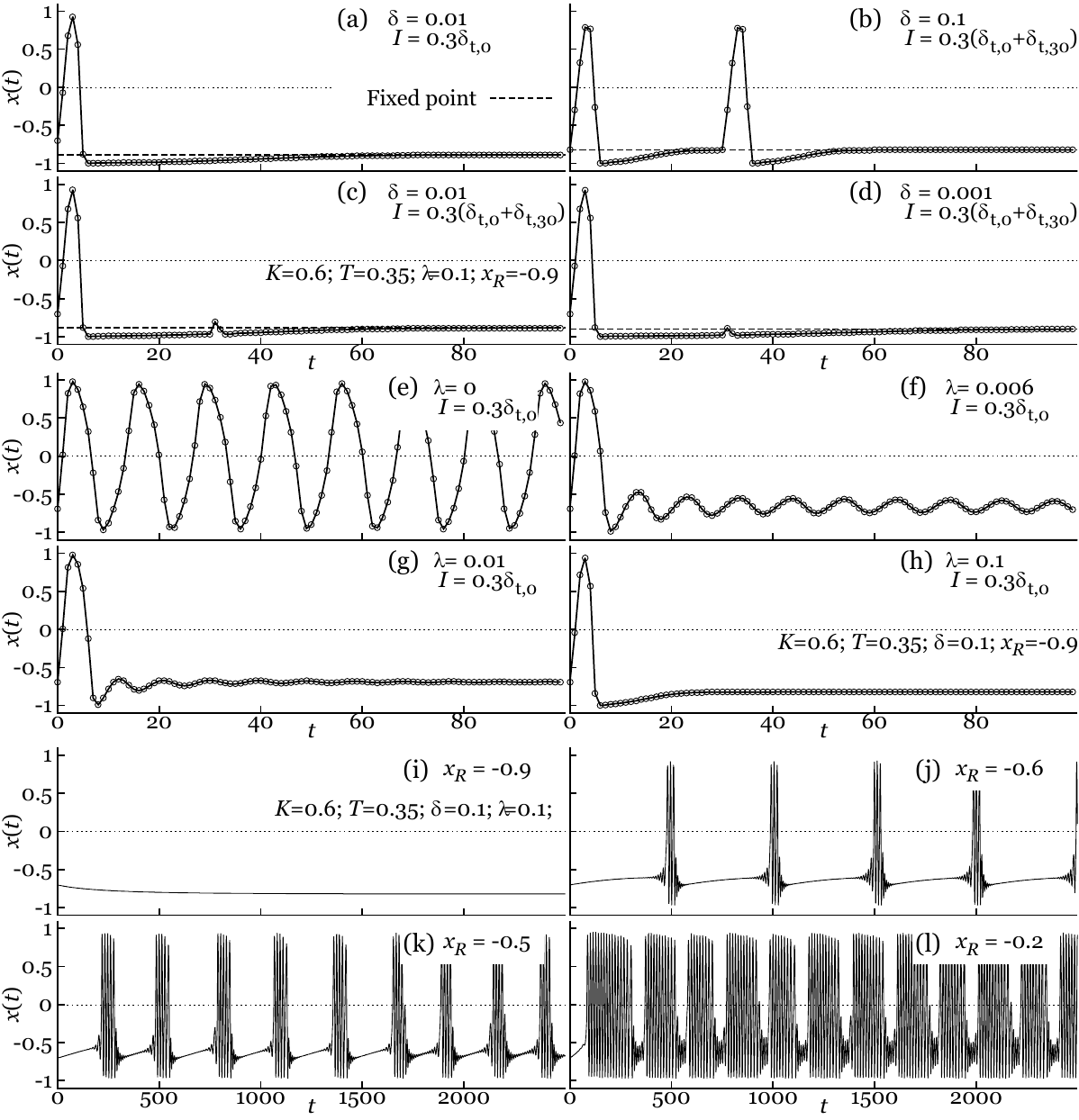}
	\end{center}
	\caption{\label{fg:KTzParam}(a) to (d): $\delta$ control the length of the refractory period; (e) to (h): $\lambda$ control the damping of oscillations; (i) to (l): $x_R$ control the bursting dynamics. Parameters' values are listed in the panels.}
\end{figure}

\citet{modeloKTz2001} modified the $z(t)$ dynamics (from Eq. \ref{eq:KTeqForz}) of the three variables KT model into a simpler, yet burster, version of the neuron, which we now
name the KTz model. The equations are:
\begin{equation}
\label{eq:KTzModel}
\left.\begin{array}{l}
x(t+1) = \tanh\left(\dfrac{x(t) - Ky(t)+ H + I(t)}{T}\right)\virg\\
y(t+1) = x(t)\virg\\
z(t+1) = \left(1-\delta\right)z(t)-\lambda\left(x(t)-x_R\right)\virg
\end{array}\right.
\end{equation}
where $x(t)$ is the membrane potential of the neuron, $y(t)$ is a recovery variable and $z(t)$ is the slow total ionic current. The $K$ and $T$
parameters
control the fast spiking dynamics, as explained above. The parameter $\delta$ is the inverse recovery time of $z(t)$,
$\lambda$ and $x_R$ control the slow spiking and bursting dynamics \cite{modeloKTz2001}. In practice, though, $\delta$, $\lambda$ and $x_R$ roles
are illustrated in Fig. \ref{fg:KTzParam}.

\begin{figure}[h!]
	\begin{center}
	\includegraphics[width=50mm]{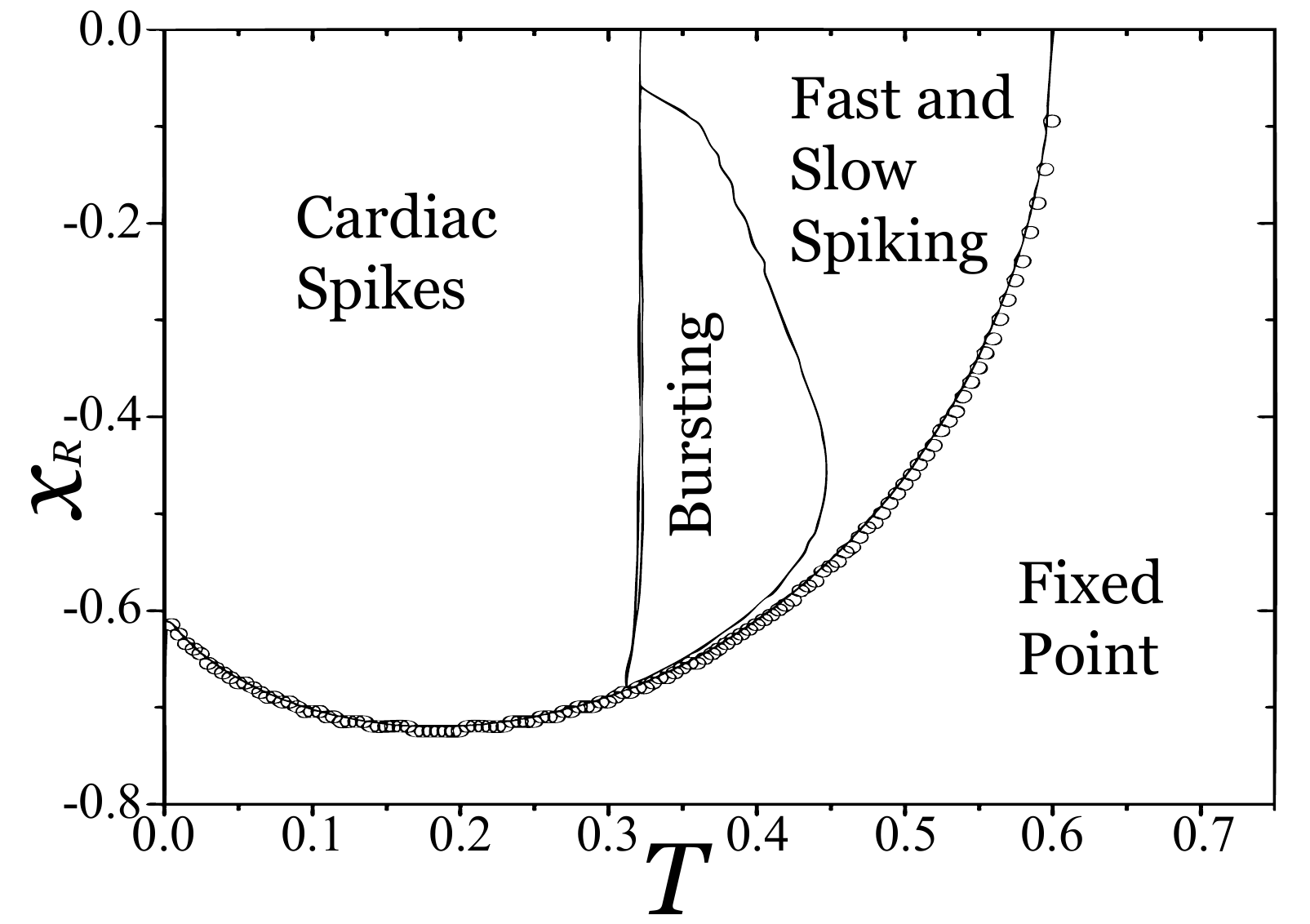}
	\end{center}
	\caption{\label{fg:KTzPhaseDiag}Detailed phase diagram for the KTz model ($\lambda=\delta=0.001$ and $K=0.6$).}
\end{figure}

\begin{figure}[h!]
	\begin{center}
	\includegraphics[width=50mm]{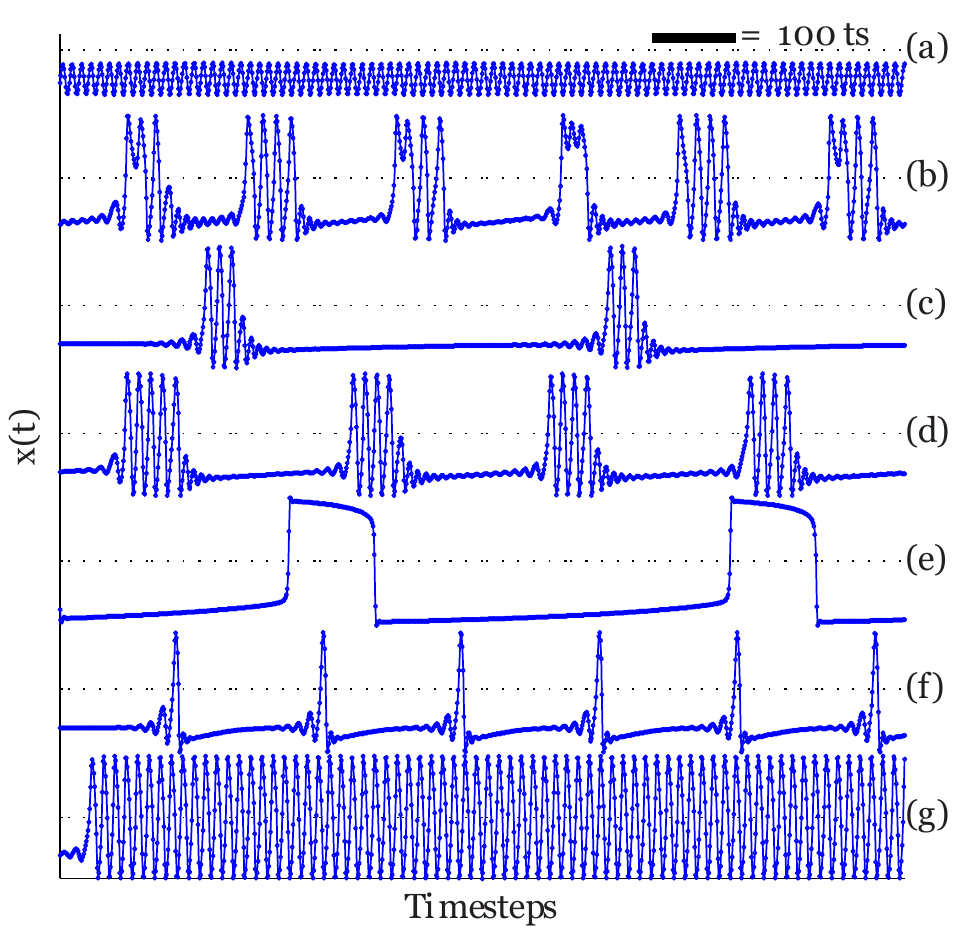}
	\end{center}
	\caption{\label{fg:KTzAut}Examples of KTz model autonomous behaviors. (a) Subthreshold oscillations; (b) Chaotic bursting; (c) Slow bursting; (d) Fast bursting; (e) Cardiac spikes; (f) Slow spiking; (g) Fast spiking.}
\end{figure}

\begin{figure}[h!]
	\begin{center}
	\includegraphics[width=80mm]{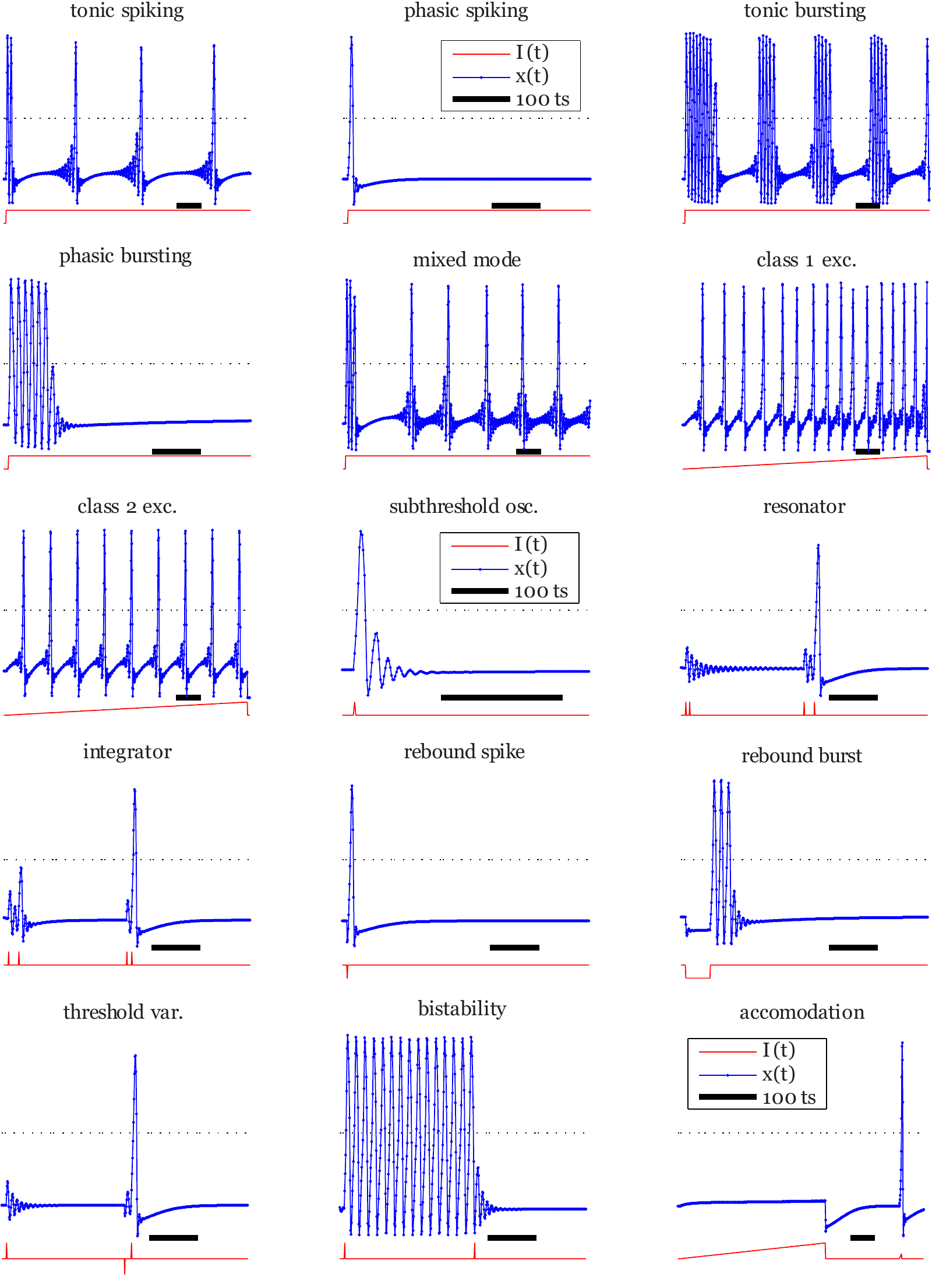}
	\end{center}
	\caption{\label{fg:KTzExc}List of the KTz model excitable behaviors. The classification of these behaviors is based on \citet{izhikevichMapas}.}
\end{figure}

KTz autonomous behaviors are listed in Fig. \ref{fg:KTzAut} whilst the excitable behaviors are in Fig. \ref{fg:KTzExc}.
A phase diagram for $\lambda=\delta=0.001$ and $K=0.6$ is given by \citet{modeloKTz2001} (see Fig. \ref{fg:KTzPhaseDiag}).
The fixed point stability of this model has been studied by
\citet{modeloKTz2004}.

We classified the KTz excitable behaviors according to \citet{izhikevichMapas}, searching for qualitative
similarities between their model and the KTz model. The latter presented $15$ out of the $20$ excitable behaviors described by
\citet{izhikevichMapas}.

Although it does not present every excitable behavior, it is a minimal model, based on only five parameters ($K$, $T$, $\delta$, $\lambda$
and $x_R$). Besides, the KTz map (Eq. \ref{eq:KTzModel}) has no singularities and the functions are not piecewise defined (as the more
recent proposals discussed in the next section). Furthermore, the great symmetry of this model facilitates its phase plane analysis, which may be accomplished through lots of detailed phase diagrams from which we can infer KTz
behavior \citep{tragtenbergYokoi,modeloKT,modeloKTz2001,modeloKTz2004}.

\subsection{Recent proposals}
\label{sec:recent}

During the last decade, three other maps have been proposed: the Rulkov model \citep{rulkovMapa}, the Izhikevich model 
citep{izhikevichMapas} and the Courbage-Nekorkin-Vdovin model \citep{courbageMap}. These and some other models have been reviewed
elsewhere \citep{courbageRev,ibarzMapas}, so we are going to present only a brief description of the first two models because
they are more easily found in literature.

\subsubsection{The Rulkov Model}
The model proposed by \cite{rulkovMapa} is, originally:
\begin{equation}
\label{eq:rulkovMap}
\left.\begin{array}{l}
	x(t+1) = F(x(t), y(t) + \beta(t))\\
	y(t+1) = y(t) - \mu(x(t)+1-\sigma(t))
\end{array}\right.\virg
\end{equation}
where $F(x,y)$ is a discontinuous function given by:
\begin{equation}
\label{eq:rulkovF}
F(x,y) = \left\{
\begin{array}{ll}
	\dfrac{\alpha}{1-x}+y & \textnormal{if } x\leq0\\
	\alpha+y & \textnormal{if } 0<x<\alpha+y\\
	-1 & \textnormal{if } x\geq \alpha+y
\end{array}
\right.\virg
\end{equation}
$\alpha$ and $\mu$ are adjustable parameters whilst $\beta(t)=\beta I(t)$ and $\sigma(t) = \sigma I(t)$ with $I(t)$ being the external input.
When there is no external input, then $\beta(t) = \beta$ and $\sigma(t) = \sigma$ become adjustable parameters.
The map consists of a fast dynamics in $x(t)$ and a slow dynamics in $y(t)$ when $\mu<<1$.

By the analysis of the nullclines of the fast and of the slow subsystems, one gets to the excitability threshold, $\sigma_{th}$,
in the $\alpha\times\sigma$ plane \cite{rulkovMapa}, which corresponds to a Subcritical Hopf Bifurcation:
\begin{equation}
\label{eq:rulkovStabLim}
\sigma_{th} = 2 - \sqrt{\alpha}\ponto
\end{equation}

\begin{figure}[h!]
	\begin{center}
	\includegraphics[width=50mm]{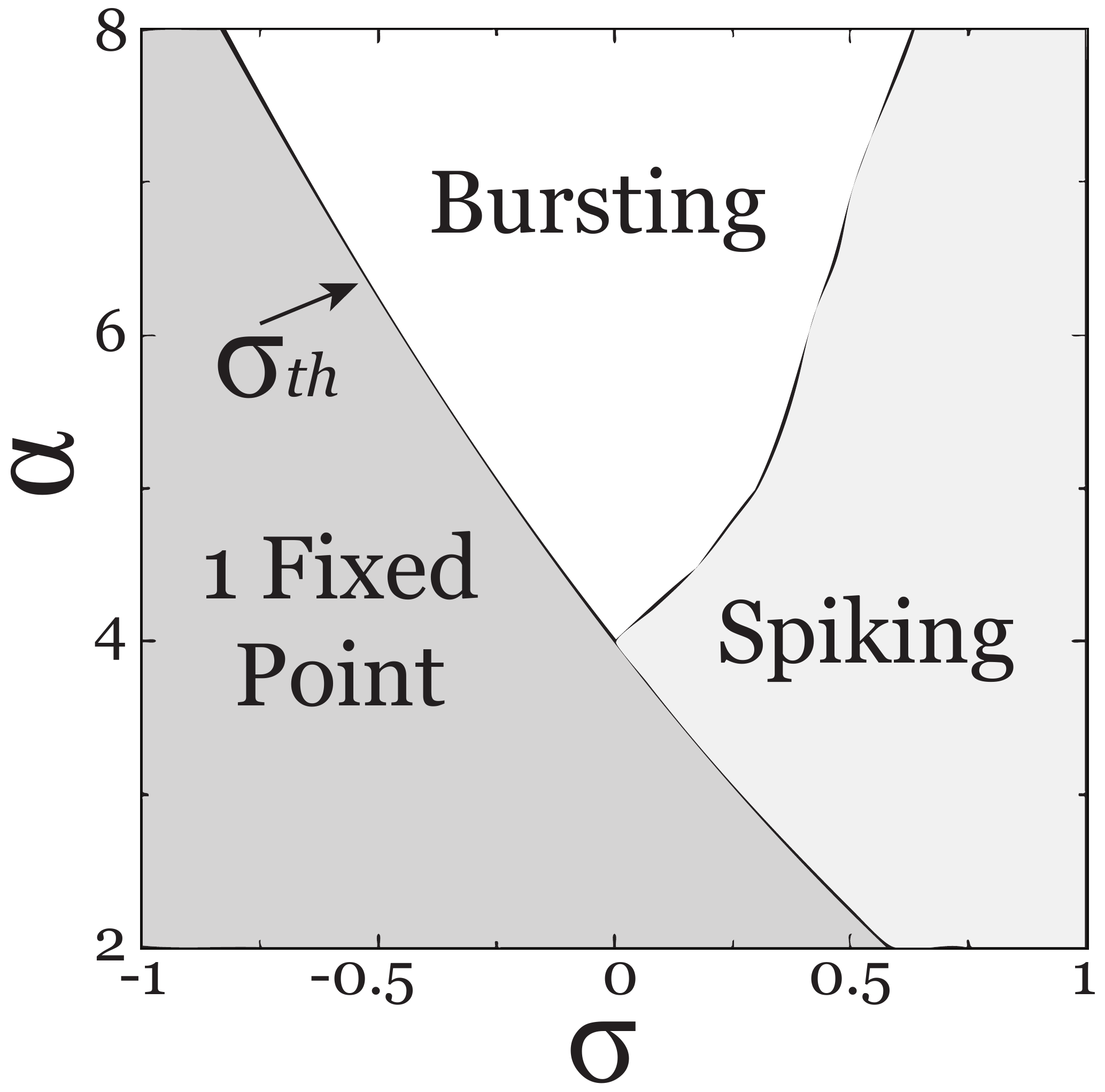}
	\end{center}
	\caption{\label{fg:rulkovPhase}Bifurcation diagram of the Rulkov model (Eq. \ref{eq:rulkovMap}). The curve $\sigma_{th}$ is given by Eq. \ref{eq:rulkovStabLim}. The stability limit between the phases of bursting and spiking is not well defined and
	has been determined computationally. Adapted from \citet{rulkovMapa}.}
\end{figure}

Eq. \ref{eq:rulkovStabLim} is plotted in Fig. \ref{fg:rulkovPhase} together with the limit between Bursting and Spiking phases
(which was determined computationally by the author). There are also a Chaotic \citep{rulkovChaotic}
and a Supercritical \citep{rulkovSupercritico} variant of the Rulkov model. The excitability of the latter is due
to a Supercritical Hopf Bifurcation, hence its name.

\subsubsection{The Izhikevich Model}
Firstly proposed as an Ordinary Differential Equation (ODE) of continous time \citep{izhikevichModel}, the map-based Izhikevich
model is the following discretization \citep{izhikevichMapas}:
\begin{equation}
\label{eq:izhikevichMap}
\left.\begin{array}{l}
x(t+1) = 0.04x^2(t)+6x(t)+140-y(t)+I(t)\\
y(t+1) = a[bx(t)-y(t)]
\end{array}\right.\virg
\end{equation}
with the reset, if $x(t)\geq30$:
\begin{equation}
\label{eq:izhReset}
\left.\begin{array}{l}
x(t+1) = c\\
y(t+1) = y(t) + d
\end{array}\right.\virg
\end{equation}
where $a$, $b$, $c$ and $d$ are the parameters, $I(t)$ is an input current (whether synaptic or
external) and $t$ is the time in mili-seconds, assumed. The parameters $a$ and $b$ control balance of the slow dynamics of the
in $y(t)$.

\begin{figure}[h!]
	\begin{center}
	\includegraphics[width=50mm]{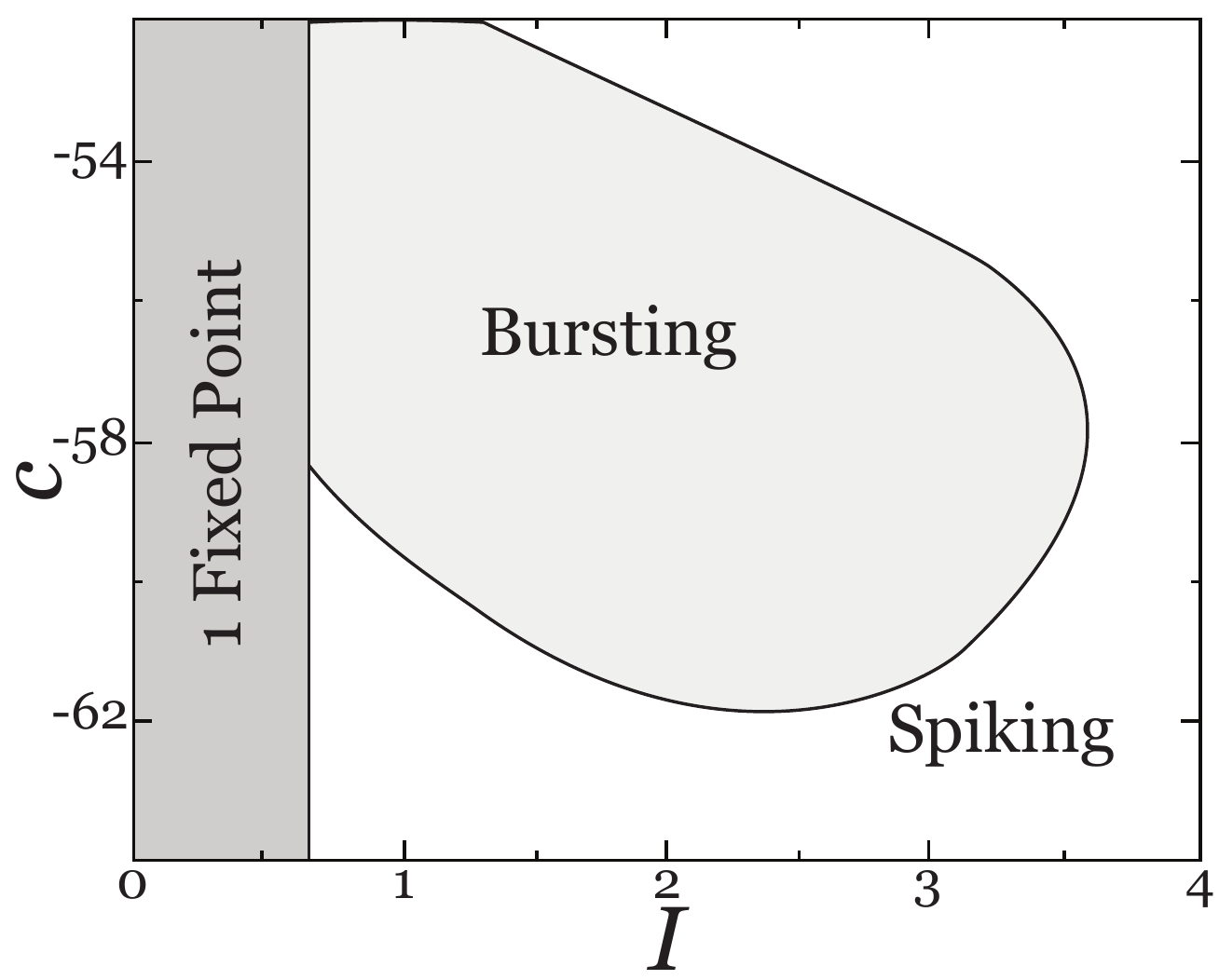}
	\end{center}
	\caption{\label{fg:izhPhase}Bifurcation diagram of the Izhikevich model (Eq. \ref{eq:izhikevichMap}) with $a=0.02$, $b=0.25$
	and $d=0$. The separation between spiking and bursting have been determined computationally and is not well defined, thus
	we traced a representative line. Adapted from \citet{ibarzMapas}.}
\end{figure}

Fig. \ref{fg:izhPhase} shows the
bifurcation diagram in the plane $c\times I$ for $a=0.02$, $b=0.25$ and $d=0$. The bifurcations undergone by the Izhikevich
model depend strongly on parameters.

The author have proposed this model as a reduced Hodgkin-Huxley model by
focusing on its bifurcations \citep{izhikevichModel}, although the equations resembles an Integrate-and-Fire (IF) model. Besides
the behaviors presented in Fig. \ref{fg:KTzExc}, this one also presents: spike frequency adaptation (similar to mixed mode panel),
spike latency (a delayed response to a stimulus), depolarizing after-potential (the membrane potential is slightly depolarized
right after the action potential), inhibitory induced spiking and inhibitory induced bursting (the onset of spiking
or bursting during the injection of an inhibitory direct current) \citep{izhikevichMapas}.

\section{Summary of the Addressed Models}
\label{sec:discus}

So far we have presented a description of the earliest maps proposed to describe neuronal
activity together with some recent approaches. \citet{izhikevichWhich} do an attempt to classify some of these models,
comparing them with the biologically motivated models (HH-like neurons). To do so, he plots a graph with
the amount of different biophysical behavior present in the model versus the amount of floating-point operations (FLOP
-- i.e. sums and multiplications)
it would take for the computer to compute $1$ ts of the model.

The Hodgkin-Huxley model may produce any biophysically plausible behavior, although it takes almost $100$ times more
FLOP. The Izhikevich model, with only $13$ FLOP, may describe $20$ different behaviors. Since the amount of FLOP is not a precise
measure of the performance of any equation (as it discard anything else than sums and multiplications, like divisions,
memory assignments, function calls, etc), we decided to do a benchmark test with the last three presented models by
averaging the CPU time, $T$ in nanoseconds (ns), each model takes to perform a total of $T_{ts}=10000$ ts. We do that
a number $S=10000\times$ to reduce the
variation of
the measures. Thus, $1$ ts takes, in average, $\bar{T}=T/T_{ts}/S$ real life nanoseconds to complete. The results are in
Table \ref{tb:modelComp} for codes running in $32$ bits and $64$ bits assemblies (both in a $64$ bits CPU). In this table,
the ellapsed time for each model, $T_M$, is calculated by $T_M=\tau T_{ts}$, where $\tau$ is the amount of miliseconds 
for each timestep of the considered model $[$ms$/$ts$]$.

\begin{center}
\begin{table}
    \begin{tabular}{l | l | l | l | l}
    \textbf{Model} & $\bar{T}_{32bit}$ $[$ns$]$ & $\bar{T}_{64bit}$ $[$ns$]$ & $\tau$ $[$ms$/$ts$]$ & $T_M$ $[$ms$]$ \\
		\hline
		\hline
        Izhikevich & $9.6\pm0.6$  & $10.7\pm0.7$ & $1^*$    & $10^4$ \\ 
        Rulkov     & $11.6\pm0.6$ & $11.5\pm0.5$ & $0.5$  & $5\times10^3$ \\
		KTz        & $30.9\pm0.9$ & $18.9\pm0.7$ & $0.1$  & $10^3$ \\
		HH         & $210\pm4$    & $255\pm5$    & $0.01$ & $10^2$
    \end{tabular}    
	\caption{\label{tb:modelComp}Comparison of CPU time, in nanoseconds $[$ns$]$, of $1$ ts for each map-based neuron model.
	We also present the CPU time of a HH-like neuron modeled by \citet{gennadyNeuro} and integrated using a 4th order Runge-Kutta method.
	All these models are adjusted in bursting
	activity. Here, $\bar{T}$ is the CPU time (real life time), $\tau$ is the conversion factor between ms and ts and $T_M$ is
	the model ellapsed time. Conversion factors are from \citet{izhikevichMapas,ibarzMapas,modeloKTz2001}. These factors
	generally take into consideration the empirical fact that $1$ spike takes, approximately, $1$ ms. $^*$ Sometimes,
	the Izhikevich conversion factor is $\tau=0.5$ ms/ts.}
\end{table}
\end{center}

Changing from $32$ bits to $64$ bits makes the KTz neuron reduce by half its CPU time, reaching the other models' performance,
whilst the others keep with their same CPU time. Nevertheless, one must remind that the choosing of the model depends strongly
on the problem one is trying to study. For example, map-based neurons are more suitable for large-scale simulations or for
generic studies (e.g. the cause of bifurcations, the rise of chaotic orbits, synchronization dynamics, avalanche dynamics,
pattern recognition, memory modeling, etc); and
biologically inspired models are more suitable for single cell studies (e.g. synaptic integration, dendritic cable filtering,
effects of dendritic morphology, the dynamics of ionic currents, etc) or even small-scale network simulations --
at least when there is worry with the simulation time.

Regardless of the chosen model and besides computational time and being a generic framework, maps bring many other advantages:
\begin{itemize}
	\item There is no need for integration nor integration timestep adjustment; In fact, adjusting an integration timestep
	turns the ODE we are trying to solve into a discretized map. The bad thing in this approach is that the stability
	of the ODE may not correspond to the map stability.
	\item The solution is exact; When solving an ODE, the adjustment of an integration timestep turns the obtained solution
	in an approximation.
	\item They are more plausible than pure cellular automaton; The latter is also used to model neuronal activity
	\citep{copelliRoqueExcMedia,copelliRoqueSensory,kinouchiCopelli,ribeiroCopelli}, however it is a very abstract
	entity in which not only the time, but even the neuronal states are discretized.
	\item They keep the main biophysical properties of the Hodgkin-Huxley-like neurons, as we have seen in the previous
	section.
\end{itemize}
Thus, one is encouraged to choose the model which fits better to his purpose: choosing the best performance may make the phase
plane analysis more difficult, as any of the Rulkov or Izhikevich model depends on a piecewise defined function. Or choosing
simplicity over performance with the KTz model.

\section{Coupling Maps}
\label{sec:coup}

The modeling of neurons as maps is an important and active field
of research in the past two decades, as we have seen in the previous sections 
of this paper. However, building up a network requires connecting
the discrete time elements through synapses distributed along
a particular network topology. This section is devoted to the
synapse models. We summarize some types of connections, then we list some 
map models for them and eventually discuss the main behaviors of these
models. We are aware of the risk of being repetitive, but we go through this
for the sake of clarity and neatness.

In short, modeling synapses may or may not take into account the types of 
connections biologically observed. We can classify the simpler map-based 
synapse models in three cathegories:
nonbiologically motivated, biologically motivated with diffusive (electrical)
character and biologically motivated with impulsive (chemical) character.  

The nonbiologically motivated models are usually rooted in the paradigm of coupled map lattices \citep{kanekoCMLbook,kanekoTsuda},
which are not necessarily neuronal networks.

Diffusive (electrical) couplings are fast connections, also called gap junctions, that couple neighbouring neurons electrically in a direct
form through channels. They are usually modelled by Ohm's law, where electric potential difference generates synaptic current through a time dependent 
(or not) conductance.

Impulsive (chemical) synapses are connections with exchange of neurotransmitters
and form the basis for neuronal communication \citep{mathPhysiol}.
Phenomenological models or biological models describe the time dependence
of the conductances, due to the release of the neurotransmitters.

In the next subsections, these three kind of models and their applications will be briefly reviewed. The neurons can also connect or disconnect as a function of time.This is called synapse plasticity and is an issue that will not
be addressed in this paper. 

\subsection{Types of Coupling}

\subsubsection{Nonbiologically Motivated Coupling}

A simplification of a neuronal network can be made if the neuron
model is disconnected from biological background and
the couplings between them are based in the paradigm of Coupled Map
Lattices, i.e., the functions are not necessarily rooted on biological foundations. 

The typical Coupled Map Lattices (CML) coupling of maps (given by Eq. \ref{eq:generalMap}) can be written in the following
generalized mathematical form \citep{ibarzMapas}:
\begin{equation}
\label{eq:CMLCoupling}
\vec{x}_i(t+1) = (1-\alpha)\vec{F}\left[\vec{x}_i(t)\right] + \dfrac{\alpha}{N_i}\sum_{j\neq i}^{N_i}\vec{F}\left[\vec{x}_j(t)\right]\virg
\end{equation}
where $\vec{x}_i(t)$ is the vector state of the $i$th neuron (node) at time $t$, with 
$1 \leq i \leq N$, $N$ being the total number of neurons, and $N_i$ is
the total number of nodes connected to the $i$th node. $\alpha\in [0;1)$ is the
coupling parameter: $\alpha=0$ means no coupling and $\alpha\rightarrow1$
means that the node $i$'s neighbors play a more important role to
this element state than the node $i$ itself.

Another kind of nonbiological coupling is to intertwin variables between elements of the network,
like did \citet{GuemesMatias} with two chaotic Chialvo neurons --
one of them drives the other -- to study chaos supression:
\begin{equation}
\label{eq:sharedCoupling}
\left.\begin{array}{l}
x_1(t+1) = [x_1(t)]^2\exp[y_1(t)-x_1(t)]+I\\
y_1(t+1) = ay_1(t) - b x_1(t) + c\\
x_2(t+1) = [x_1(t)]^2\exp[y_2(t)-x_1(t)]+I\\
y_2(t+1) = ay_2(t) - b x_2(t) + c
\end{array}\right.\ponto
\end{equation}
Notice that $x_2(t+1)$ is function of $x_1(t)$ and $y_2(t)$, and not of $x_2(t)$ as expected. Thus, it is said that the first neuron drives the second one.
The synchronization may supress chaos in a couple of Chialvo neurons 
with intertwined variables.

It is important to notice that the framework of Eq. \ref{eq:CMLCoupling}
does not hold if one chooses to work with more elaborate neuron models,
like the KTz or the Rulkov models. The phase plane analysis and the
bifurcation diagrams of such models provide a very specific way of
inputing external currents into their equations, via the variable $I_i(t)$.

The so-called pulse-coupled neural network (PCNN) is an example of simple
coupling through input currents. The input over neuron $i$ is assumed to be:
\begin{equation}
\label{eq:pulseCoupling}
I_i(t) = \sum_{j\neq i}^{N_i}J_{ij}x_j(t)\virg
\end{equation}
where $J_{ij}$ is the coupling intensity (the conductance of the channel, in the case of a synapse). The sum runs over the 
$N_i$ neuron $i$'s neighbors and the $x_j(t)$'s are the presynaptic
membrane potentials. It is called \textit{pulse coupling}
because, as soon as the neuron $j$ starts a pulse, it is readily transmitted to neuron $i$ scaled by the intensity $J_{ij}$ in the next time step.

As examples of PCNN, take the \citet{modeloKT} approach to study
emergence of collective oscillatory state: a fully connected network of KT neurons (Sec. \ref{sec:ktz}) coupled via Eq. \ref{eq:pulseCoupling}
homogeneously ($J_{ij}=J_{ji}=J$) and heterogeneously. The coupled model reads:
\begin{equation}
\label{eq:KTCoupled}
\left.\begin{array}{l}
x_i(t+1) = \tanh\left[\dfrac{x_i(t)-Ky_i(t)+H+\sum_{j\neq i}^{N_i}J_{ij}x_j(t)}{T}\right]\\
y_i(t+1) = x_i(t)
\end{array}\right.\ponto
\end{equation}

Another example is the chaotic \citet{rulkovChaotic} model in a mean field pulse-coupled network, implemented by the equation:
\begin{equation}
\label{eq:rulkovPulseCoup}
\left.\begin{array}{l}
x_i(t+1)= \dfrac{\alpha_i}{1 + [x_i(t)]^2} + y_i(t) + \dfrac{\epsilon}{N} \sum\limits_{j\neq i}^N x_j(t)\\
y_i(t+1)=y_i(t) - \mu[x_i(t)-\sigma_i]
\end{array}\right.\virg
\end{equation}
used to study the synchronization of bursts when the network is not homogeneous (notice the indices in $\sigma_i$ and in $\alpha_i$). In this case, pulse coupling
allows chaotic bursts to synchronize.

The next two kinds of coupling are biologically inspired and, hence, they are modeled by input currents just like the one
used by PCNN.

\subsubsection{Diffusive Coupling}
\label{sec:diff}

Generally speaking, diffusive (electrical) couplings, also called gap junctions,
 are fast couplings, where channels between neighboring cells are formed and 
allow ions or small molecules to pass through them. Their exchange speed allow faster synchronization
than chemical synapses. They are usual in the heart, other muscles
 and during development where synchronization play a major role \citep{mathPhysiol}, but in
vertebrates they are a minority.

Typically, they have the form \citep{gapJunctionRef}:
\begin{equation}
\label{eq:GapJunction}
I_{i}(t) = \sum\limits_{j\neq i}^{N_i}J_{ij}[x_j(t)-x_i(t)]\virg
\end{equation}
where $x_{i,j}$ are membrane potentials and $J_{ij}$ is the conductance of the gap junction channel.

\subsubsection{Impulsive Coupling}
\label{sec:imp}

The couplings via neurotransmitter exchanges are slower than the electrical ones,
but more frequent in vertebrates. They are also called chemical of impulsive couplings
and are responsible by the signal processing.

Generally, the synaptic current $I_{syn}(t)$ is given by
\begin{equation}
\label{eq:ChemicalCoupling}
I_{syn}(t) = g_{syn} (t) [x_{post}(t) - E_{syn}]
\end{equation}

where $g_{syn} (t)$ is the synaptic current, $x_{post}(t)$ the postsynaptic 
membrane potential, $E_{syn}$ is
the reversal postsynaptic potential. The synaptic conductance may be modeled
by a instantaneous rise with single exponential decay, an alpha function (with a continuous 
rise and fall) and a difference of exponentials \citep{synapticModelSchutter}. 

Another punch line is the fast threshold modulation (FTM) approximation \citep{Somers1993}, where there is a very sharp activation threshold and a constant conductance for each coupling such that
\begin{equation}
\label{eq:FTMCoupling}
I_i(t) = \sum_k I_{i,k}(t)
\end{equation}
where
\begin{equation}
\label{eq:FTMCurrent}
I_{i,k}(t) = g_{k} [x_i(t) - x_{r,k} ]\sum_k  H[x_j(t)-\theta_k] 
\end{equation}
where H(x) is the step function, and the index k labels the different types of chemical synapses.
The synaptic reversal potential is $x_{r,k}$ and $\theta_k$ are the presynaptic threshold of 
activation.

However, there is indeed a time scale (or more) for the synaptic coupling. \citet{RulkovBazhenov2004} and \citet{BazhenovRulkov2005}
propose an equation for the synaptic current:
\begin{equation}
\label{eq:RulkovCoupling}
I_{syn} (t+1) =  \gamma I_{syn} (t) - g[x_{post}-x_r] \delta(t-t_{pre,k})
\end{equation}
where g is a conductance and $t_{pre,k}$ is the time steps the presynaptic neuron
has fired.

%%%%%%
%%%%%%
%%%%%% INICIO MODELO KUVA
%%%%%%
%%%%%%
\citet{modeloKTz2001} proposed a map-based equation to model the chemical synaptic coupling between neurons. Their objective is to allow
further research on Computational Neuroscience via biologically motivated Coupled Map Lattices.
The input over neuron $i$ due to coupling current is the simple sum of the synaptic currents:
\begin{equation}
\label{eq:KuvaCurrent}
I_i(t) = \sum_{j}^{N_i}Y_{ij}(t)\virg
\end{equation}
where $N_i$ is the amount of neighbors of neuron $i$ and $Y_{ij}(t)$ is the synaptic current given by:
\begin{equation}
\label{eq:KuvaSyn}
\left.\begin{array}{l}
	Y_{ij}(t+1) = \left(1-\dfrac{1}{\tau_1}\right)Y_{ij}(t) + h_{ij}(t)\\
	h_{ij}(t+1) = \left(1-\dfrac{1}{\tau_2}\right)h_{ij}(t) + J_{ij}\Theta[x_j(t)]
\end{array}\right.\virg
\end{equation}
where $\tau_1$ and $\tau_2$ are exponential time constants, $J_{ij}$ is the coupling intensity (with dimension of conductance)
and $\Theta(x)=1$ if $x>0$ and $0$ otherwise is the Heaviside function. This synaptic current may be excitatory ($J>0$) or inhibitory ($J<0$).

\begin{figure}[h!]
	\begin{center}
	\includegraphics[width=50mm]{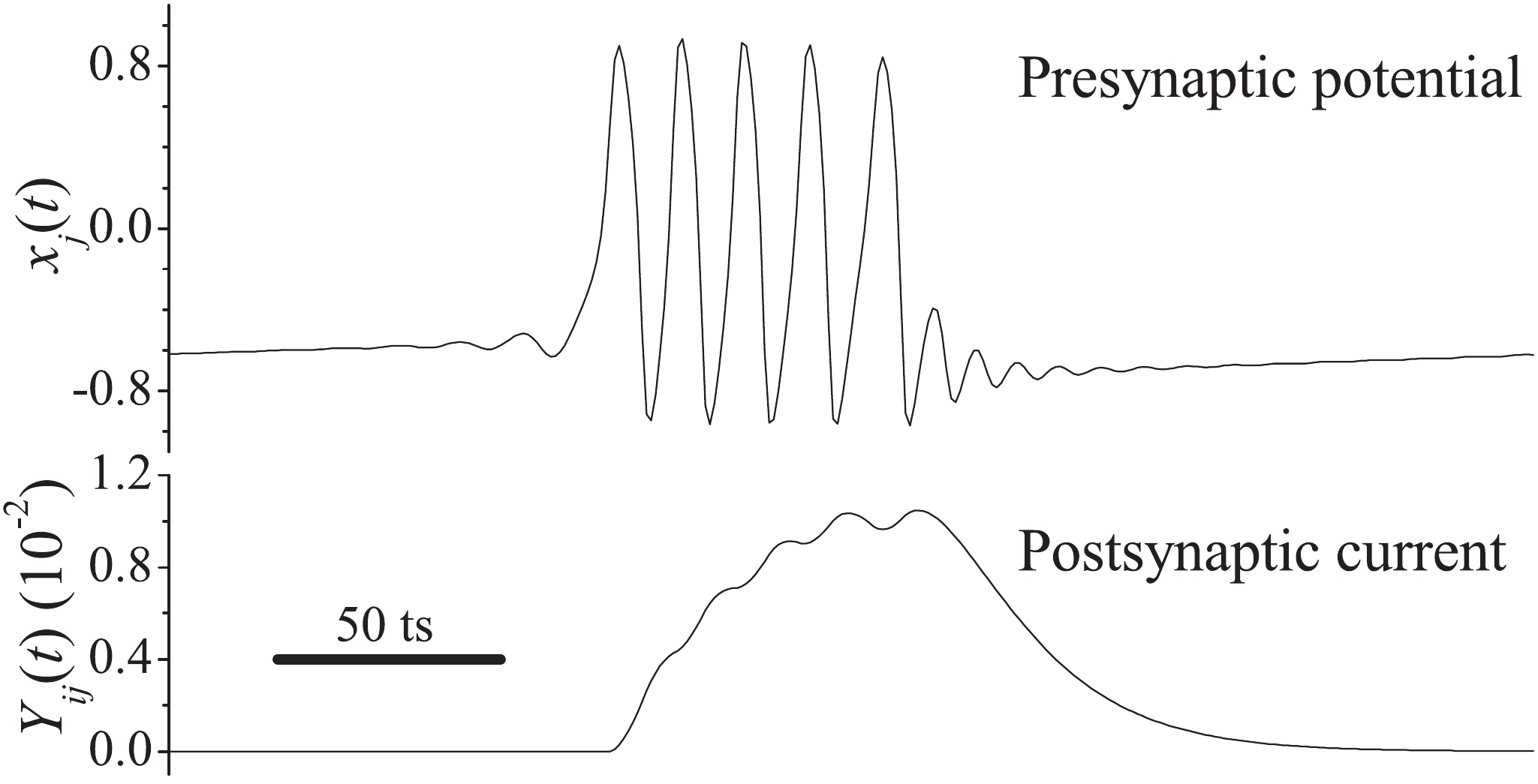}
	\end{center}
	\caption{\label{fg:KuvaSyn}Time evolution of the postsynaptic current, Eq. \ref{eq:KuvaSyn} for $J=0.0001$,
	$\tau_1=\tau_2=15$ (bottom panel) due to a presynaptic KTz neuron in bursting regime, Eq. \ref{eq:KTzModel} with $K=0.6$, $T=0.35$, $\delta=\lambda=0.001$ and $x_R=-0.5$ (top panel).}
\end{figure}

So far, notice that Eq. \ref{eq:KuvaSyn} is an approximation of a double exponential function,
$f(t)=C[\exp(-t/\tau_1)-\exp(-t/\tau_2)]$, which reduces to an alpha function, $f(t)=Ct\exp(-t/\tau)$,
when $\tau_1=\tau_2=\tau$; $C$ is a constant -- see Fig. \ref{fg:KuvaSyn}. Generally, this equation is used to model the conductance of synaptic currents.
However, it is still a good approximation for modeling synaptic currents \citep{deSchutterBook}.

Moreover one may question why there is no coupling parameter multiplying the synaptic current, $Y_{ij}(t)$, in Eq. \ref{eq:KuvaCurrent}.
The coupling parameter is $J$. If one wish to multiply a new constant, say $C$, in that term: $CY_{ij}(t)$, then a simple change of variables in
Eq. \ref{eq:KuvaSyn} would vanish with $C$, keeping the same dynamical behavior:
$Y'_{ij}(t)\rightarrow CY_{ij}(t)$, $h'_{ij}(t)\rightarrow Ch_{ij}(t)$ and $J'\rightarrow CJ$.
%%%%%%
%%%%%%
%%%%%% FIM MODELO KUVA
%%%%%%
%%%%%%

\begin{figure}[h!]
	\begin{center}
	\includegraphics[width=50mm]{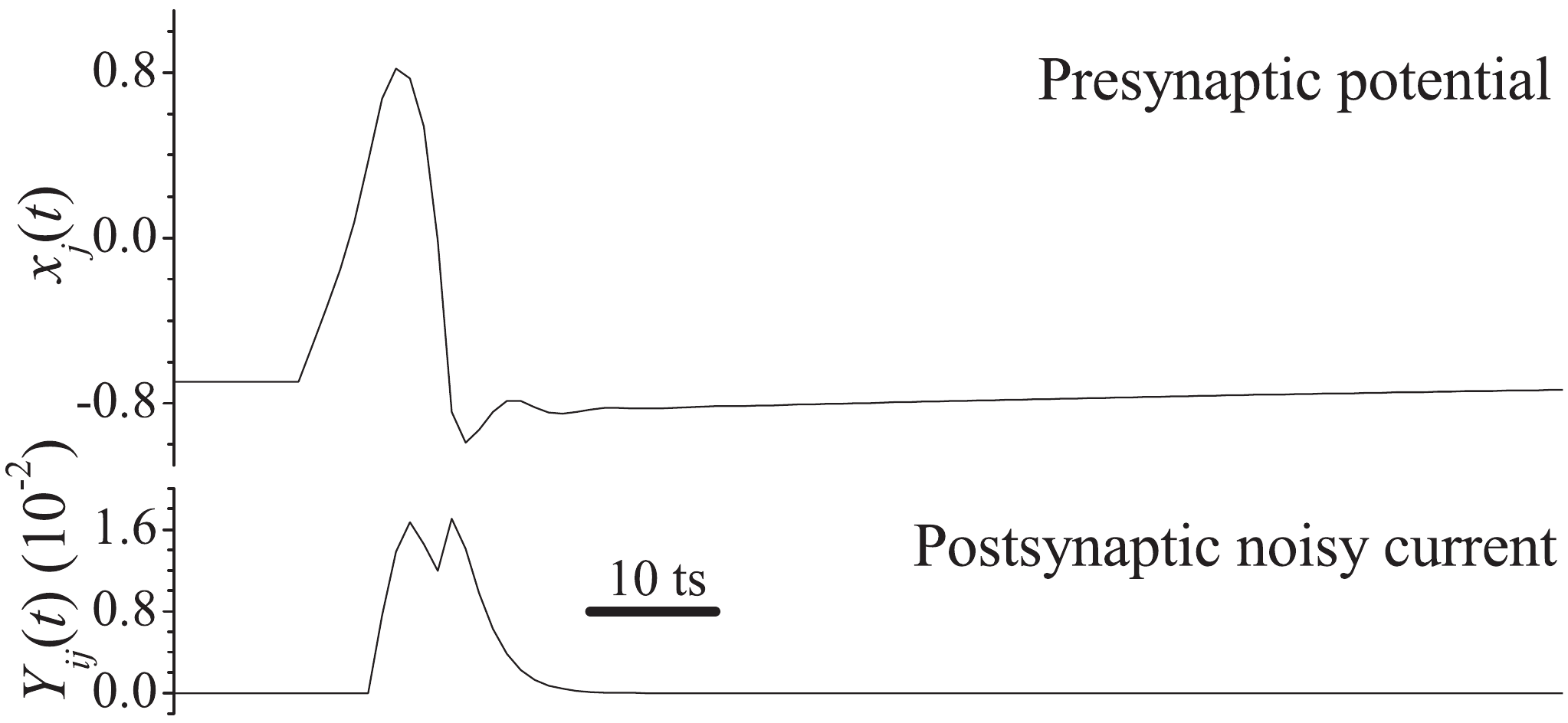}
	\end{center}
	\caption{\label{fg:GKTSyn}Time evolution of the postsynaptic noisy current, Eq. \ref{eq:KuvaSyn} with $J_{ij}(t)$ given by
	Eq. \ref{eq:GKTCoupling} with $J=0.001$, $R=0.01$ and
	$\tau_1=\tau_2=2$ (bottom panel) due to a presynaptic KTz neuron spike, Eq. \ref{eq:KTzModel} with $K=0.6$, $T=0.35$, $\delta=0.001$, $\lambda=0.008$ and $x_R=-0.7$ (top panel).}
\end{figure}

%%%%%%
%%%%%%
%%%%%% INICIO MODELO GKT
%%%%%%
%%%%%%
\citet{girardiAva} studied a complete map-based neural network by using KTz maps coupled with Kuva synapses.
The authors also adapted the Kuva
synaptic current (Eq. \ref{eq:KuvaSyn}), by adding a uniform random noise term in the coupling, in order to model
synaptic noise observed experimentally \citep{peretto}. Then, the Girardi-Schappo-Kinouchi-Tragtenberg (GKT) model is composed of
Eq. \ref{eq:KuvaSyn} with $J_{ij}\equiv J_{ij}(t)$ given by:
\begin{equation}
\label{eq:GKTCoupling}
J_{ij}(t) = J + \epsilon_{ij}(t)\virg
\end{equation}
where $J$ is the coupling parameter and $\epsilon_{ij}(t)$ is the uniform random time signal of amplitude $|R|$. The noise is different for every synapse $j\rightarrow i$.
To keep the synaptic coherence (i.e. in order to the keep inhibitory synapses always inhibitting and excitatory synapses always exciting), the sign of $R$ is the same
sign of $J$, then
if $J>0$, we have $\epsilon_{ij}(t) \in [0;R]$, otherwise we have $\epsilon_{ij}(t) \in [-R;0]$. The noisy synaptic current is plotted in Fig. \ref{fg:GKTSyn}.

The important feature of this model is that the synapse is not always effective. It has a probability $q$ of exciting the postsynaptic neuron. Similar
to the input current threshold, above which a neuron fires, there is a threshold in the coupling parameter, $J_{th}$, above which the postsynaptic neuron will fire
(the threshold $J_{th}$ may also be negative when the postsynaptic neuron is excitable by rebound).

A first glance into Eq. \ref{eq:GKTCoupling} lead us to define the probability, $p$, of having $|J_{ij}(t)|>|J_{th}|$:
\begin{equation}
\label{eq:GKTThreshold}
p = \dfrac{J + R - J_{th}}{R}\ponto
\end{equation}
Note that $|J+R|>|J_{th}|$ always, because $J$, $R$ and $J_{th}$ have always the same sign. One could expect that $q=p$. That is not true because there is
an internal dynamics in the synaptic current equations: as long as the spike takes place ($x_j(t)>0$), the current keeps summing different quantities
$J+\epsilon_{ij}(t)$. Thus, the same set $(J,R)$ may not always lead to a postsynaptic spike. Anyway, the quantity $p$ may be regarded as a first approximation to $q$.
%%%%%%
%%%%%%
%%%%%% FIM MODELO GKT
%%%%%%
%%%%%%

\subsection{Applications of the synapse models}

A summary of the results obtained by each of the classes of
synapse models (nonbiologically motivated, electric, chemical)
in recognizing image patterns, modeling cat visual cortex, 
synchronization (connected with topology network, noise, spike-burst behavior, delay,
rewiring, ...), bistable states of the brain, effects of external signals, information transmission and criticality in neuronal avalanches is given in this subsection.

\subsubsection{Nonbiologically motivated}

%This section is devoted to the applications of two approaches: PCNN, in particular, and CML, in general.

As one example of CML application is
the study of synchonization made by \citet{Jampa}. They connected Chialvo neurons through a short version of Eq. \ref{eq:CMLCoupling}, in which
the coupling term is written only as $(\epsilon/2)[x_{j(t)}(t) + x_{k(t)}(t)]$, where the indices $j(t)$ and $k(t)$ are picked at each timestep,
with probability $p$, by
randomly assigning to each node an index, $l\in[1;N]$, and then calculating $j(t)=l-1$ and $k(t)=l+1$ (modulo $N$). The rewiring probability $p$ and the coupling strength 
$\epsilon$ determine the kind of synchronization: chaotic, spatio-temporal chaos or fixed point. 

On the other hand, the PCNN approach is used for many purposes:
 modeling cat visual cortex\citep{eckhornModel}, 
recognizing image patterns \citep{pulseRev}, synchronization and related problems
\citep{modeloKT,ermentroutMapPulse,zouMapPulse}.

\citet{eckhornModel} modeled the cat visual cortex via PCNN, particularly the stimulus-induced oscillations, used for pattern recognition and feature associations.
\citet{pulseRev} reviewed the successful use of PCNN in many dimensions of image processing: segmentation, denoising, object and edge detection, feature extraction and pattern recognition (enhancement, fusion, etc).
\citet{ermentroutMapPulse} discretized an ODE, and found waves in a ring and synchonization stability dependent of the number of neurons.
Stable chaos was found in a PCNN of discontinuous maps by \citet{zouMapPulse}.

\citet{Pontes} found that the coupling strength needed to synchronized Rulkov bursters
is smaller when the range of couplings is bigger.

\citet{Vries2001a, Vries2001b} showed that mean field-coupling of the same system
can turn isolated spiking into coupled bursting neurons. Then, in this case
bursting can be seen as an emergent behavior.

\citet{Politi2012} used coupled logistic maps, as well as Stuart-Landau oscillators and leaky IF neurons, to study dynamical properties of coupled oscillator sparse lattices in many scales of length. They show a transition from micro to macroscopic scale in network syncronization driven by a critical conectivity.

 \citet{LopezRuiz2012} mimicked a bistable (sleep-awaken) brain by a logistic map lattice, with usual CML interactions.

%%%%
%%%%
%%%% FEITO POR MAURICIO
%%%%
%%%%

The pulse-coupled KT model with homogeneous couplings, $J_{ij}=J$, 
may lead a network of silent neurons (adjusted with $(K,T,H<H_s)$ --
Eq. \ref{eq:KTStabLim}, outside of the bubble in Fig. \ref{fg:KTPhaseDiag} into a collective oscillating phase if the sum
$I_i(t)=J\sum_{\lmean j\rmean}x_j(t)$ is sufficiently big -- such that $\tilde{H} = H + I_i(t)>H_s(K,T)$ (for neurons with $H<0$).

More clearly, if $H<0$, the fixed point is also $x^*<0$. Thus, if $J<0$ (inhibitory coupling), the sum
$I_i(t)=J\sum_{\lmean j\rmean}x_j(t)$ will be a positive number and the achieved state will have a new $\tilde{H}$ given by
$\tilde{H}=H+I_i(t)$.
Then, there is a threshold -- exactly equal to the excitability threshold discussed in Sec. \ref{sec:ktz} -- above which the
network will spontaneously oscillate, even though any uncoupled neuron is silent (in the fixed point).
This is a counter-intuitive result, although it also happens
with realistic neurons \citep{rinzelPulse,ernstPulse}.

If the term $|J|$ is big enough, the network will cross through the bubble in Fig. \ref{fg:KTPhaseDiag} and will
reach the upper fixed point, $x^*>0$. This picture may be generalized to the heterogeneous case, however the value of $I_i(t)$, in
such case, is not guaranteed to be positive.

The heterogeneous case may be used both for modeling practical problems and for investigation purposes. As examples, one could assemble
a neural network for pattern recognition by mixing linear neurons (in the output) and KT units and adjusting weights according to a training algorithm \citep{modeloKT} (see \citet{albanoNN} for an application
with neurons given by Eq. \ref{eq:MPModel}, but with a nonlinear activation function).
On the other hand, \citet{izhikevichModel} studies brain rythms via a PCNN made of his neuron model (discretized with a $0.1$ ms timestep).

The Izhikevich heterogeneous network consists in mixed excitatory ($J_{ij}>0$) and inhibitory ($J_{ij}<0$) coupling.
From the network's time evolution, emerge groups of synchronized neurons in different phases --
the so called \textit{polychronous} state. \citet{izhikevichPoly} discuss that these polychronous groups represent
the memory of the network, as different external stimuli produce different polychronization patterns.
Yet, the number of produced patterns is far greater than the number of neurons in the network, yielding a very big memory
capacity.

%%%% deixa a parte abaixo ou tira?
%%%% deixa a parte acima ou tira?
%
%%%%
%%%%
%%%% FIM FEITO POR MAURICIO
%%%%
%%%%

\subsubsection{Diffusive coupling}

%%%% ORGANIZADO POR MAURICIO
%%%%
%%%%
Diffusive coupling has been extensively used in the study of synchronization, spatio-temporal chaos, effects of external inputs on a neuronal network, spiking-bursting transition, among other subjects.

\citet{rulkovMapa} studied the existence of synchronization regimes for spiking and bursting activities of rulkov-map a function of  neuron coupling strength.

\citet{WeiLuo2007} connected Rulkov neurons in a small-world network
 through  typical electrical connections to study synchronization and spatio-temporal chaos. 

Subthreshold stimulus were found to induce synchronization in a noisy square lattice
Rulkov network by \citet{Wang2007}. The study of  Rulkov neurons with information transmission delay in a small-world geometry with additive noise and delay, as done by \citet{Wang2008}, show the existence of transitions in synchronization as a function of the delay. The delay may also induce stochastic resonance in a
scale-free network of Rulkov neurons, have shown \citet{Wang2009}.

The increasing of diversity of Rulkov neuron connections may induce
synchronization in a Rulkov map neural network, as analyzed by \citet{Chen2008}.
The influence of a mix of electrical and chemical synapses connecting Rulkov neurons on burst synchonization and transmission (seen as chaotic itinerancy \citet{kanekoChItinerancy})
were studied by \citet{Tanaka2006}. \citet{Ivanchenko2007} and  \citet{Ivanchenko2004} studied transitions between bursting and spiking Rulkov neurons connected by electrical synapses in many time scales and the chaotic phase synchronization by an external periodic input to a single neuron.
\citet{Batista2007} and \citet{Batista2009} also showed that external periodic signals can generate burst synchronization in networks of Rulkov neurons electrically coupled.

\subsubsection{Impulsive coupling}
Connecting map neurons by chemical synapses is not as popular as using electrical synapses,
despite the chemical are more widespread than the electrical. One of the most popular
approaches is the FTM, which shows no synaptic dynamics. However, there are some proposals
of describing the chemical synapse dynamics applied to network behavior.

In the FTM front, \citet{Ivanchenko2007} showed that electrical and impulsive (chemical)
synapses behave like electrical, since networks of Rulkov maps exhibit transition syncronized-desynchronized transition as a function of the coupling strength.

\citet{Ibarz2008} used FTM to show a bunch of effects of impulsive synapses: excitatory
synapses may generate antiphase synchronization, synapse may change from excitatory to
inhibitory as a change in conductance and with the same reversal potential, small
variations in the synaptic threshold may cause big changes in synchronization of
spikes within bursts.

\citet{ShiLu2009} also applied FTM coupling in the studiyg of in-phase burst
synchronization of Rulkov  neurons. 

Inhibitory bursting Rulkov maps chemically coupled may exhibit
power-law behavior at the onset of a synchronization transition, driven by the coupling strength or stimulation current\citep{Franovic2010}.

The effect of delay in in-phase and anti-phase synchronization was the object of \citet{Franovic2011} in Rulkov map neural network connected by reciprocal sigmoid chemical synapses.

%%%% FEITO POR MAURICIO
%%%%
%%%%

Another different approach is to consider connections with their own dynamics.
\citet{RulkovBazhenov2004} proposed a chemical synapse model, given by
\begin{equation}
I_{syn}(t+1) = \gamma I_{syn}(t) - g_{syn}[x_{post}(t) - x_r]\delta(t-t_{pre,k})
\end{equation}
where the delta function is 1 for a presynaptic spike and 0 otherwise. $\gamma$ is
a decay constant, for an single exponential dynamics in synapse. The model exhibit
many dynamic behaviors and is qualitatively comparable to Hodgkin-Huxley model.
Using basically the same model, \citet{BazhenovRulkov2005} determined
resonance properties on collective behavior in a cortical network model containing excitatory
 and inhibitory cells and showed that network interactions can enhance the frequency 
range of reliable responses, such that the latter can be controlled
by the strength of synaptic connections.

Dynamical map models for impulsive synapses were also proposed by \citet{modeloKTz2001} and 
\citet{girardiAva}. The first one uses uniform synapses and the second introduces synaptic
noise, believed to be crucial to information processing. With their model, \citet{girardiAva} could study the effects of the synaptic inefficacy makes a network of neurons coupled with GKT synapses a matter of stochastic dynamics, since the amount of neurons, $s$, that
will fire in the network due to an initial stimulation is not obvious. The network activity due to one stimulus is called an avalanche.
The authors investigated regular square lattices with free boundary conditions of KTz neurons coupled with GKT synapses and
found that $s$ follows a powerlaw $P(s)\sim s^{-\mu}$ with $\mu=1.35$. Besides, the duration of avalanches, $w$, also follows
a powerlaw $P(w)\sim w^{-\nu}$, with $\nu=1.50$.

Moreover, the authors measured the activity of a subsampled network -- a network in which the measurement can be made
in only a very small fraction of the elements. Generally, experimental setups are subsampled, since
the number of neurons which are recorded is far less than the number of neurons in the sample \citep{violaSub,ribeiroCopelli}.
The subsampling results matched the experiments of \citet{ribeiroCopelli} and, together with the powerlaw behavior in the network activity,
both in temporal and spatial dimensions, 
show that the system is critical and suggests that there may be a Self-Organized Critical (SOC) state in brain activity and neural network models \citep{girardiAva}.

The SOC state is, generally, present when there is a balance between tensioning and relaxing the system
\citep[and many others]{bakPRL,bakPRA,christensenFFM,jensenSOC,dharSOCformalismo,bonachela1}. Due to theoretical considerations about
criticality in neural networks \citep{linkenkaerRef141,stassiBrain,herzCML}, \citet{beggsPlenz2003} proposed SOC to explain the brain activity.
Avalanches are experimentally observed
\citep[and references there in]{beggsPlenz2003,beggsPlenz2004,violaSub,ribeiroCopelli,chialvoReview,wernerFractais,beggsCritical,plenzBenefits},
although there is a great debate on the criticality of the brain.

The basic idea is that the brain works with the precisely needed amount of activity and that there may be
a homeostatic mechanism which led the brain to such a state. Within this framework, mental diseases and disorders would be associated with deviations
from the critical amount of activity \citep{chialvoCritical,chialvoReview,vertesCNS}.
The critical state has been shown to enhance the dynamical range of sensitivity to external stimuly \citep{kinouchiCopelli,plenzDynRange}, to optimize the memory and learning
processes \citep{socPlasticity} and the computational power of the brain \citep{wernerFractais,plenzBenefits,beggsCritical}.

A necessary condition for SOC is that the system may not be
imposed into the critical state, rather it should find its own path towards criticality with no fine tuning. As no computer model
could work on its own, without external tuning of parameters, \citet{kinouchiQuasiSOC} proposed to call these systems Self-Organized quasi-Critical (SOqC). SOqC
systems have a balance dynamical equation that makes the system oscillate in the neighborhood of the critical point.
In this sense, the GKT model is not SOC, nor SOqC, because the authors had to adjust the parameters $J$ and $R$. However, a variation of the GKT model could contain
a homeostatic mechanism over the parameter $R$ or $J$, for instance, that would take the system towards the critical point, turning the system into a SOqC system.
%%%%
%%%%
%%%% FIM FEITO POR MAURICIO
%%%%
%%%%

\section{Concluding Remarks}
\label{sec:conc}

Although generally studied in the context of dynamical systems, some of the models outlined in this work may be used in computational applications, such as
pattern recognition, data analysis, data classification, data association and so on, like the McCulloch-Pitts model utilized in the perceptron.
We directed our studies mainly through the
historical development of the map-based models, by theoretically constructing over the McCulloch-Pitts model, adding delayed self-couplings (which creates dynamical
behaviors), changing to a continuous activation function and adding coupling currents to the model.

This approach allowed us to close the gap between the first discrete time models and the most recent maps. In general, we tried to keep the original
design concepts of the models, except for the Izhikevich's, which was originally designed as an ODE.
Nevertheless we assume that the discretization of an ODE results in a map,
with behavior of its own. Thus, we linked two entire family of models with very rich excitable and dynamical behaviors,
namely the perceptron family (from McCulloch-Pitts until KTz) and the HH family (from Chialvo to Izhikevich). We also showed that there is a close dynamical
correspondence between KTz and the behaviors of HH.

The models of the HH family are, generally, piecewise defined functions. However, all of these models present bursting activity with
only two dynamical variables, differently from the KTz map (whose
bursting behavior was based in Hindmarsh-Rose model). On the other hand, the symmetry and simplicity of the KTz map may be of great help during the phase plane
and bifurcation analysis. KT and KTz also may be suitable when there is a need to explore the effects of the whole action potential, as all the other models
assume that the spike occurs instantly. Therefore, if one wants to model a more rigorous synaptic transmission with the map framework, KT or KTz are should be chosen.

Every map model presents a similar computational performance, as discussed in Sec. \ref{sec:discus}, which is approximately $20\times$ faster than the HH performance.
Thus, the most prominent features of all the studied models are their simiplicity, reliability, numerical stability and computational performance.

Moreover, we categorized the many types of coupling used to connect the covered models. Many times, the coupling and the network topology are indissociable, forcing
one to study the coupling of neurons under a given topology. From pulse-coupled networks to map-based modeling of chemical synapses, we presented their most common usability,
relating with the results achieved in each case. Since synapses is, alone, a broad research area, we highlight the importance of modeling synapses specifically for working
with map-based neural networks, as did \citet{modeloKTz2001}, otherwise the differential equations would have to be discretized for each case, bringing additional modeling
problems, namely how to synchronize the time evolution of the map-based neuron with the time evolution of the synapses, as the latter should be precisely solved to
avoid numerical inaccuracy?

Finally, recent results have shown that map-based neural networks is a promising developing field, for example,
\citet{girardiAva} modeled both synapses and neurons
with maps and showed that a critical state may develop in regular lattices when the coupling is subjected to uniform noise.
Just as examples, on the list of some still unexplored paths in map-based neuronal modeling is the compartimental modeling and
the seek for a \textit{canonical} purely map-based model (in the ODE framework, the neuron model differs from the whole
class of HH-type neurons by just a change o variables \citep{izhikevichCanon};
on the other hand, a canonical map-based model would be the most
simple model capable of describing all the HH-type neurons behaviors).

With this work, we hope we have brought more
attention to this kind of modeling, which may play an important role in the forthcoming years, both in technologic applications and in neuroscientific research.

\section{Acknowledgement}
We would like to thank the invitation to make this neuron map modeling overview from Antonio Carlos Roque da Silva.

\bibliographystyle{elsarticle-harv}

\begin{thebibliography}{103}
\expandafter\ifx\csname natexlab\endcsname\relax\def\natexlab#1{#1}\fi
\expandafter\ifx\csname url\endcsname\relax
  \def\url#1{\texttt{#1}}\fi
\expandafter\ifx\csname urlprefix\endcsname\relax\def\urlprefix{URL }\fi

\bibitem[{Aihara and Suzuki(2010)}]{aihara2010}
Aihara, K., Suzuki, H., 2010. Theory of hybrid dynamical systems and its
  applications to biological and medical systems. Phil. Trans. R. Soc. A 368,
  4893--4914.

\bibitem[{Aihara et~al.(1990)Aihara, Takabe, and Toyoda}]{Aihara1990}
Aihara, K., Takabe, T., Toyoda, M., 1990. Chaotic neural networks. Phys. Lett.
  A 144(6,7), 333--340.

\bibitem[{Albano et~al.(1992)Albano, Passamante, Hediger, and
  Farrell}]{albanoNN}
Albano, A.~M., Passamante, A., Hediger, T., Farrell, M.~E., 1992. Using neural
  nets to look for chaos. Physica D 58, 1--9.

\bibitem[{Azevedo et~al.(2009)Azevedo, Carvalho, Grinberg, Farfel, Ferretti,
  Leite, Filho, Lent, and Herculano-Houzel}]{suzanaBrain}
Azevedo, F. A.~C., Carvalho, L. R.~B., Grinberg, L.~T., Farfel, J.~M.,
  Ferretti, R. E.~L., Leite, R. E.~P., Filho, W.~J., Lent, R.,
  Herculano-Houzel, S., 2009. Equal numbers of neuronal and nonneuronal cells
  make the human brain an isometrically scaled-up primate brain. J. Comp.
  Neurol. 513, 532--541.

\bibitem[{Bak et~al.(1987)Bak, Tang, and Wiesenfeld}]{bakPRL}
Bak, P., Tang, C., Wiesenfeld, K., 1987. Self-organized criticality: An
  explanation of 1/f noise. Phys. Rev. Lett. 59(4), 381--384.

\bibitem[{Bak et~al.(1988)Bak, Tang, and Wiesenfeld}]{bakPRA}
Bak, P., Tang, C., Wiesenfeld, K., 1988. Self-organized criticality. Phys. Rev.
  A 38(1), 364--374.

\bibitem[{Batista et~al.(2009)Batista, Batista, de~Pontes, Lopes, and
  Viana}]{Batista2009}
Batista, C. A.~S., Batista, A.~M., de~Pontes, J. A.~C., Lopes, S.~R., Viana,
  R.~L., 2009. Bursting synchronization in scale-free networks. Chaos, Solitons
  and Fractals 41, 2220--2225.

\bibitem[{Batista et~al.(2007)Batista, Batista, de~Pontes, Viana, and
  Lopes}]{Batista2007}
Batista, C. A.~S., Batista, A.~M., de~Pontes, J. A.~C., Viana, R.~L., Lopes,
  S.~R., 2007. Bursting synchronization in scale-free networks. Phys. Rev. E
  76, 016218.

\bibitem[{Bazhenov et~al.(2005)Bazhenov, Rulkov, Fellous, and
  Timofeev}]{BazhenovRulkov2005}
Bazhenov, M., Rulkov, N.~F., Fellous, J., Timofeev, I., 2005. Role of network
  dynamics in shaping spike timing reliability. Phys. Rev. E 72, 041903.

\bibitem[{Beggs and Plenz(2003)}]{beggsPlenz2003}
Beggs, J.~M., Plenz, D., 2003. Neuronal avalanches in neocortical circuits. J.
  Neurosci. 23(35), 11167--11177.

\bibitem[{Beggs and Plenz(2004)}]{beggsPlenz2004}
Beggs, J.~M., Plenz, D., 2004. Neuronal avalanches are diverse and precise
  activity patterns that are stable for many hours in cortical slice cultures.
  J. Neurosci. 24(22), 5216--5229.

\bibitem[{Beggs and Timme(2012)}]{beggsCritical}
Beggs, J.~M., Timme, N., 2012. Being critical of criticality in the brain.
  Front. Physiol. 3, 163.

\bibitem[{Bellman(2003)}]{curseDim}
Bellman, R.~E., 2003. Dynamic Programming. Dover.

\bibitem[{Bonachela and Muñoz(2009)}]{bonachela1}
Bonachela, J.~A., Muñoz, M.~A., 2009. Self-organization without conservation:
  true or just apparent scale-invariance? J. Stat. Mech., P09009.

\bibitem[{Bower and Beeman(2003)}]{genesisBook}
Bower, J.~M., Beeman, D., 2003. The Book of GENESIS: Exploring Realistic Neural
  Models with the GEneral NEural SImulation System. Internet Edition.

\bibitem[{Caianello(1961)}]{caianello1961}
Caianello, E.~R., 1961. Outline of a theory of thought process and thinking
  machines. J. Theor. Biol. 1, 204--235.

\bibitem[{Carnevale and Hines(2006)}]{neuronBook}
Carnevale, N.~T., Hines, M.~L., 2006. The NEURON Book. Cambridge University
  Press.

\bibitem[{Chen and Liu(2009)}]{Chen2008}
Chen, H., Liu, J. Z.~J., 2009. Enhancement of neuronal coherence by diversity
  in couple rulkov-map models. Physica A 19, 023112.

\bibitem[{Chialvo(1995)}]{chialvo1995}
Chialvo, D.~R., 1995. Generic excitable dynamics on a two-dimensional map.
  Chaos Solitons Fractals 5, 461--479.

\bibitem[{Chialvo(2004)}]{chialvoCritical}
Chialvo, D.~R., 2004. Critical brain networks. Physica A 340, 756--765.

\bibitem[{Chialvo(2010)}]{chialvoReview}
Chialvo, D.~R., 2010. Emergent complex neural dynamics. Nat. Phys. 6, 744--750.

\bibitem[{Christensen et~al.(1993)Christensen, Flyvbjerg, and
  Olami}]{christensenFFM}
Christensen, K., Flyvbjerg, H., Olami, Z., 1993. Self-organized critical
  forest-fire model: Mean-field theory and simulation results in 1 to 6
  dimensions. Phys. Rev. Lett. 71(17), 2737--2740.

\bibitem[{Connors and Long(2004)}]{gapJunctionRef}
Connors, B.~W., Long, M.~A., 2004. Electrical synapses in the mammalian brain.
  Annual Review of Neuroscience 27, 393--418.

\bibitem[{Copelli et~al.(2005)Copelli, Oliveira, Roque, and
  Kinouchi}]{copelliRoqueSensory}
Copelli, M., Oliveira, R.~F., Roque, A.~C., Kinouchi, O., 2005. Signal
  compression in the sensory periphery. Neurocomputing 65--66, 691--696.

\bibitem[{Copelli et~al.(2002)Copelli, Roque, Oliveira, and
  Kinouchi}]{copelliRoqueExcMedia}
Copelli, M., Roque, A.~C., Oliveira, R.~F., Kinouchi, O., 2002. Physics of
  psychophysics: Stevens and weber-fechner laws are transfer functions of
  excitable media. Phys. Rev. E 65, 060901.

\bibitem[{Copelli et~al.(2004)Copelli, Tragtenberg, and
  Kinouchi}]{modeloKTz2004}
Copelli, M., Tragtenberg, M. H.~R., Kinouchi, O., 2004. Stability diagrams for
  bursting neurons modeled by three-variable maps. Physica A 342, 263--269.

\bibitem[{Courbage and Nekorkin(2009)}]{courbageRev}
Courbage, M., Nekorkin, V.~I., 2009. Map based models in neurodynamics. Int. J.
  Bifurcat. Chaos 20(6), 1631--1651.

\bibitem[{Courbage et~al.(2007)Courbage, Nekorkin, and Vdovin}]{courbageMap}
Courbage, M., Nekorkin, V.~I., Vdovin, L.~V., 2007. Chaotic oscillations in a
  map-based model of neural activity. Chaos 17(4), 043109.

\bibitem[{Cymbalyuk and Calabrese(2000)}]{gennadyNeuro}
Cymbalyuk, G., Calabrese, R.~L., 2000. Oscillatory behaviors in
  pharmacologically isolated heart interneurons from the medicinal leech.
  Neurocomputing 32--33, 97--104.

\bibitem[{Dayan and Abbott(2001)}]{abbottNeuro}
Dayan, P., Abbott, L.~F., 2001. Theoretical Neuroscience: Computational and
  Mathematical Modeling of Neural Systems. The MIT Press.

\bibitem[{de~Arcangelis et~al.(2006)de~Arcangelis, Perrone-Capano, and
  Herrmann}]{socPlasticity}
de~Arcangelis, L., Perrone-Capano, C., Herrmann, H.~J., 2006. Self-organized
  criticality model for brain plasticity. Phys. Rev. Lett. 96, 028107.

\bibitem[{de~Schutter(2010)}]{deSchutterBook}
de~Schutter, E. (Ed.), 2010. Computational Modeling Methods for Neurocientists.
  The MIT Press.

\bibitem[{de~Vries(2001{\natexlab{a}})}]{Vries2001a}
de~Vries, G., 2001{\natexlab{a}}. Bursting as an emergent phenomenon in coupled
  chaotic maps. Phys. Rev. E 64, 051914.

\bibitem[{de~Vries(2001{\natexlab{b}})}]{Vries2001b}
de~Vries, G., 2001{\natexlab{b}}. From spikers to bursters via coupling: help
  from heterogeneity. Bull. Math. Biol 63, 371--391.

\bibitem[{de~Vries(2012{\natexlab{a}})}]{Politi2012}
de~Vries, G., 2012{\natexlab{a}}. Collective dynamics in sparse networks. Phys.
  Rev. Lett. 109, 138103.

\bibitem[{de~Vries(2012{\natexlab{b}})}]{WeiLuo2007}
de~Vries, G., 2012{\natexlab{b}}. Collective dynamics in sparse networks. Phys.
  Rev. Lett. 109, 138103.

\bibitem[{Dhar(2006)}]{dharSOCformalismo}
Dhar, D., 2006. Theoretical studies of self-organized criticality. Physica A
  369, 29--70.

\bibitem[{Eckhorn et~al.(1990)Eckhorn, Reitboeck, Arndt, and
  Dicke}]{eckhornModel}
Eckhorn, R., Reitboeck, H.~J., Arndt, M., Dicke, P.~W., 1990. Feature linking
  via synchronization among distributed assemblies: simulations of results on
  cat visual cortex. Neural Computation 2, 293--307.

\bibitem[{Ernst et~al.(1995)Ernst, Pawelzik, and Geisel}]{ernstPulse}
Ernst, U., Pawelzik, K., Geisel, T., 1995. Synchronization induced by temporal
  delays in pulse-coupled oscillators. Phys. Rev. Lett. 74, 1570--1573.

\bibitem[{FitzHugh(1955)}]{fitzhugh}
FitzHugh, R., 1955. Mathematical models of threshold phenomena in the nerve
  membrane. Bulletin of Mathematical Biophysics 17, 257--278.

\bibitem[{Franovic and Miljkovic(2010)}]{Franovic2010}
Franovic, I., Miljkovic, V., 2010. Power law behavior related to mutual
  synchronization of chemically coupled map neurons. Euro. Phys. J. B 76,
  613--624.

\bibitem[{Franovic and Miljkovic(2011)}]{Franovic2011}
Franovic, I., Miljkovic, V., 2011. The effects of synaptic time delay on motifs
  of chemically coupled rulkov model neurons. Commun. Nonlinear Sci. Numer.
  Simul. 16, 623--633.

\bibitem[{Girardi-Schappo et~al.(2012)Girardi-Schappo, Kinouchi, and
  Tragtenberg}]{girardiAva}
Girardi-Schappo, M., Kinouchi, O., Tragtenberg, M. H.~R., 2012. Critical
  avalanches and subsampling in map-based neural networks. arXiv:1209.3271
  [cond-mat.dis-nn].

\bibitem[{Goel and Ermentrout(2002)}]{ermentroutMapPulse}
Goel, P., Ermentrout, B., 2002. Synchrony, stability, and firing patterns in
  pulse-coupled oscillators. Physica D 163, 191--216.

\bibitem[{Golomb and Rinzel(1993)}]{rinzelPulse}
Golomb, D., Rinzel, J., 1993. Dynamics of globally coupled inhibitory neurons
  with heterogeneity. Phys. Rev. E 48, 4810--4814.

\bibitem[{G\"u\'emez and Mat\'ias(1996)}]{GuemesMatias}
G\"u\'emez, J., Mat\'ias, M.~A., 1996. Synchronous oscillatory activity in
  assemblies of chaotic model neuron. Physica D 96, 334--343.

\bibitem[{Herz and Hopfield(1995)}]{herzCML}
Herz, A. V.~M., Hopfield, J.~J., 1995. Earthquake cycles and neural
  reverberations: Collective oscillations in systems with pulse-coupled
  threshold elements. Phys. Rev. Lett. 75, 1222--1225.

\bibitem[{Hindmarsh and Rose(1984)}]{modeloHR}
Hindmarsh, J.~L., Rose, R.~M., 1984. A model of neuronal bursting using three
  coupled first order differential equations. Proc. R. Soc. Lond., B, Biol.
  Sci. 221, 87--102.

\bibitem[{Hopfield(1984)}]{Hopfield1984}
Hopfield, J.~J., 1984. Neurons with graded response have collective
  computational properties like those of two-sate neurons. Proc. Nat. Acad.
  Sci. (USA) 81, 3088--3092.

\bibitem[{Hoppensteadt and Izhikevich(2002)}]{izhikevichCanon}
Hoppensteadt, F.~C., Izhikevich, E.~M., 2002. Canonical Neural Models. The MIT
  Press.

\bibitem[{Ibarz et~al.(2008)Ibarz, Cao, and Sanjuan}]{Ibarz2008}
Ibarz, B., Cao, H., Sanjuan, M. A.~F., 2008. Bursting regimes in map-based
  neuron models coupled through fast threshold modulation. Phys. Rev. E 77,
  051918.

\bibitem[{Ibarz et~al.(2011)Ibarz, Casado, and Sanjuán}]{ibarzMapas}
Ibarz, B., Casado, J.~M., Sanjuán, M. A.~F., 2011. Map-based models in neuronal
  dynamics. Phys. Rep. 501, 1--74.

\bibitem[{Ivanchenko et~al.(2004)Ivanchenko, Osipov, Shalfeev, and
  Kurths}]{Ivanchenko2004}
Ivanchenko, M.~V., Osipov, G.~V., Shalfeev, V.~D., Kurths, J., 2004. Phase
  synchronization in ensembles of bursting oscillators. Phys. Rev. Lett. 93,
  134101.

\bibitem[{Ivanchenko et~al.(2007)Ivanchenko, Osipov, Shalfeev, and
  Kurths}]{Ivanchenko2007}
Ivanchenko, M.~V., Osipov, G.~V., Shalfeev, V.~D., Kurths, J., 2007. Network
  mechanism for burst generation. Phys. Rev. Lett. 98, 108101.

\bibitem[{Izhikevich(2003)}]{izhikevichModel}
Izhikevich, E.~M., 2003. Simple model of spiking neurons. IEEE Trans. Neural
  Netw. 14(6), 1569--1572.

\bibitem[{Izhikevich(2004)}]{izhikevichWhich}
Izhikevich, E.~M., 2004. Which model to use for cortical spiking neurons? IEEE
  Trans. Neural Netw. 15, 1063--1070.

\bibitem[{Izhikevich(2006)}]{izhikevichPoly}
Izhikevich, E.~M., 2006. Polychronization: Computation with spikes. Neural
  Comput. 18(2), 245--282.

\bibitem[{Izhikevich and Hoppensteadt(2004)}]{izhikevichMapas}
Izhikevich, E.~M., Hoppensteadt, F., 2004. Classification of bursting mappings.
  Int. J. Bifurcat. Chaos 14(11), 3847--3854.

\bibitem[{Jampa et~al.(2007)Jampa, Sonawane, Gade, and Sinha}]{Jampa}
Jampa, M. P.~K., Sonawane, A.~R., Gade, P.~M., Sinha, S., 2007. Sincronization
  in a network of model neurons. Physical Review E 75, 026215.

\bibitem[{Jensen(1998)}]{jensenSOC}
Jensen, H.~J., 1998. Self-Organized Criticality: Emergent Complex Behavior in
  Phyical and Biological Systems. Cambridge University Press.

\bibitem[{Kaneko(1993)}]{kanekoCMLbook}
Kaneko, K., 1993. Theory and Applications of Coupled Map Lattices. Wiley.

\bibitem[{Kaneko(1994)}]{kanekoCML}
Kaneko, K., 1994. Relevance of dynamic clustering to biological networks.
  Physica D 75, 55--73.

\bibitem[{Kaneko and Tsuda(2001)}]{kanekoTsuda}
Kaneko, K., Tsuda, I., 2001. Complex Systems: Chaos and Beyond. A Constructive
  Approach with Applications in Life Sciences. Springer.

\bibitem[{Kaneko and Tsuda(2003)}]{kanekoChItinerancy}
Kaneko, K., Tsuda, I., 2003. Chaotic itinerancy. Chaos 13, 926--936.

\bibitem[{Keener and Sneyd(1998)}]{mathPhysiol}
Keener, J., Sneyd, J., 1998. Mathematical Physiology. Springer.

\bibitem[{Kinouchi(1998)}]{kinouchiQuasiSOC}
Kinouchi, O., Fevereiro 1998. Self-organized (quasi-)criticality: the extremal
  feder and feder model. arXiv:cond-mat/9802311v1.

\bibitem[{Kinouchi and Copelli(2006)}]{kinouchiCopelli}
Kinouchi, O., Copelli, M., 2006. Optimal dynamical range of excitable networks
  at criticality. Nat. Phys. 2, 348--351.

\bibitem[{Kinouchi and Tragtenberg(1996)}]{modeloKT}
Kinouchi, O., Tragtenberg, M. H.~R., 1996. Modeling neurons by simple maps.
  Int. J. Bifurcat. Chaos 6, 2343--2360.

\bibitem[{Kuva et~al.(2001)Kuva, Lima, Kinouchi, Tragtenberg, and
  Roque}]{modeloKTz2001}
Kuva, S.~M., Lima, G.~F., Kinouchi, O., Tragtenberg, M. H.~R., Roque, A.~C.,
  2001. A minimal model for excitable and bursting elements. Neurocomputing
  38--40, 255--261.

\bibitem[{Little(1974)}]{little1974}
Little, W.~A., 1974. The existence of persistent states in the brain. Math.
  Biosci. 19, 101--120.

\bibitem[{López-Ruiz and Fournier-Prunaret(2012)}]{LopezRuiz2012}
López-Ruiz, R., Fournier-Prunaret, D., August 2012. The bistable brain: a
  neuronal model with symbiotic interactions. arXiv:nlin.CD/1208.0223v1.

\bibitem[{McCulloch and Pitts(1943)}]{McCulloch1943}
McCulloch, W.~S., Pitts, W.~H., 1943. A logical calculus of the ideas immanent
  in nervous activity. Bull. Math. Biophys. 5, 115--133.

\bibitem[{Morris and Lecar(1981)}]{morrisLecar}
Morris, C., Lecar, H., 1981. Voltage oscillations in the barnacle giant muscle
  fiber. Biophysics Journal 35, 193--213.

\bibitem[{Nagumo et~al.(1962)Nagumo, Arimoto, and Yoshizawa}]{nagumo}
Nagumo, J., Arimoto, S., Yoshizawa, S., 1962. An active pulse transmission line
  simulating nerve axon. Proceedings of the IRE 50, 2061--2070.

\bibitem[{Nagumo and Sato(1972)}]{nagumo1972}
Nagumo, J., Sato, S., 1972. On a response characteristic of a mathematical
  neuron model. Kybernetik 10, 155--164.

\bibitem[{Pasemann(1993)}]{pasemann1993}
Pasemann, F., 1993. Dynamics of a single model neuron. Int. J. Bifurcat. Chaos
  2, 271--278.

\bibitem[{Pasemann(1997)}]{pasemann1997}
Pasemann, F., 1997. A simple chaotic neuron. Physica D 104, 205--2011.

\bibitem[{Peretto(1994)}]{peretto}
Peretto, P., 1994. An Introduction to the Modeling of Neural Networks.
  Cambridge University Press.

\bibitem[{Pontes et~al.(2008)Pontes, Viana, Lopes, Batista, and
  Batista}]{Pontes}
Pontes, J. C.~A., Viana, R.~L., Lopes, S.~R., Batista, C. A.~S., Batista,
  A.~M., 2008. Bursting synchronization in non-locally coupled maps. Physica A
  387, 4417--4428.

\bibitem[{Priesemann et~al.(2009)Priesemann, Munk, and Wibral}]{violaSub}
Priesemann, V., Munk, M. H.~J., Wibral, M., 2009. Subsampling effects in
  neuronal avalanche distributions recorded in vivo. BMC Neurosci. 10, 40.

\bibitem[{Ribeiro et~al.(2010)Ribeiro, Copelli, Caixeta, Belchior, Chialvo,
  Nicolelis, and Ribeiro}]{ribeiroCopelli}
Ribeiro, T.~L., Copelli, M., Caixeta, F., Belchior, H., Chialvo, D.~R.,
  Nicolelis, M. A.~L., Ribeiro, S., 2010. Spike avalanches exhibit universal
  dynamics across the sleep-wake cycle. PLoS ONE 5(11), e14129.

\bibitem[{Roth and van Rossum(2010)}]{synapticModelSchutter}
Roth, A., van Rossum, M. C.~W., 2010. Modeling Synapses. The MIT Press.

\bibitem[{Rulkov(2001)}]{rulkovChaotic}
Rulkov, N.~F., 2001. Regularization of synchronized chaotic bursts. Phys. Rev.
  Lett. 86, 183--186.

\bibitem[{Rulkov(2002)}]{rulkovMapa}
Rulkov, N.~F., 2002. Modeling of spiking-bursting neural behavior using
  two-dimensional map. Phys. Rev. E 65, 041922.

\bibitem[{Rulkov et~al.(2004)Rulkov, Timofeev, and
  Bazhenov}]{RulkovBazhenov2004}
Rulkov, N.~F., Timofeev, I., Bazhenov, M., 2004. Oscillations in large-scale
  cortical networks: map-based model. J. Comput. Neurosci. 17, 203--223.

\bibitem[{Shew and Plenz(2013)}]{plenzBenefits}
Shew, W.~L., Plenz, D., 2013. The functional benefits of criticality in the
  cortex. Neuroscientist 19(1), 88--100.

\bibitem[{Shew et~al.(2009)Shew, Yang, Petermann, Roy, and
  Plenz}]{plenzDynRange}
Shew, W.~L., Yang, H., Petermann, T., Roy, R., Plenz, D., 2009. Neuronal
  avalanches imply maximum dynamic range in cortical networks at criticality.
  J. Neurosci. 29(49), 15595--15600.

\bibitem[{Shi and Lu(2009)}]{ShiLu2009}
Shi, X., Lu, Q., 2009. Burst synchronization of electrically and chemically
  coupled map neurons. Physica A 388, 2410--2419.

\bibitem[{Shilnikov and Rulkov(2004)}]{rulkovSupercritico}
Shilnikov, A.~L., Rulkov, N.~F., 2004. Subthreshold oscillations in a map-based
  neuron model. Phys. Lett. A 328, 177--184.

\bibitem[{Somers and Kopell(1993)}]{Somers1993}
Somers, D., Kopell, N., 1993. Rapid synchronization through fast threshold
  modulation. Biol. Cybern. 68, 393--407.

\bibitem[{Stassinopoulos and Bak(1995)}]{stassiBrain}
Stassinopoulos, D., Bak, P., 1995. Democratic reinforcement: A principle for
  brain function. Phys. Rev. E 51(5), 5033--5039.

\bibitem[{Tanaka et~al.(2006)Tanaka, Ibarz, Sanjuán, and Aihara}]{Tanaka2006}
Tanaka, G., Ibarz, B., Sanjuán, M. A.~F., Aihara, K., 2006. Synchronization and
  propagation of bursts in networks of coupled map neurons. Chaos 16, 013113.

\bibitem[{Tragtenberg and Yokoi(1995)}]{tragtenbergYokoi}
Tragtenberg, M. H.~R., Yokoi, C. S.~O., 1995. Field behavior of an ising model
  with competing interactions on the bethe lattice. Phys. Rev. E 52(3),
  2187--2197.

\bibitem[{Usher et~al.(1995)Usher, Stemmler, and Olami}]{linkenkaerRef141}
Usher, M., Stemmler, M., Olami, Z., 1995. Dynamic pattern formation leads to
  1/f noise in neural populations. Phys. Rev. Lett. 74, 326--329.

\bibitem[{Veiga and Tragtenberg(2001)}]{modeloKTRessonancia}
Veiga, F. L.~S., Tragtenberg, M. H.~R., 2001. A very stochastic resonant neuron
  model. Neurocomputing 38--40, 423--431.

\bibitem[{Vertes et~al.(2011)Vertes, Bassett, and Duke}]{vertesCNS}
Vertes, P.~E., Bassett, D.~S., Duke, T., 2011. Scale-free statistics of
  neuronal assemblies predict learning performance. BMC Neurosci. 12(Suppl 1),
  O4.

\bibitem[{Wang et~al.(2008)Wang, Duan, Perc, and Chen}]{Wang2008}
Wang, Q.~Y., Duan, Z., Perc, M., Chen, G.~R., 2008. Synchonization transitions
  on small-world neuronal networks: effects of information transmission delay
  and rewiring probability. Europhys. Lett. 83, 50008.

\bibitem[{Wang et~al.(2007)Wang, Lu, and Chen}]{Wang2007}
Wang, Q.~Y., Lu, Q.~S., Chen, G.~R., 2007. Subthreshold stimulus-aided temporal
  order and synchronization in a square lattice noisy neuronal network.
  Europhys. Lett. 77, 10004.

\bibitem[{Wang et~al.(2009)Wang, Perc, Duan, and Chen}]{Wang2009}
Wang, Q.~Y., Perc, M., Duan, Z., Chen, G.~R., 2009. Delay-induced multiple
  stochastic resonances on scale-free neuronal networks. Chaos 19, 023112.

\bibitem[{Wang et~al.(2010)Wang, Ma, Cheng, and Yang}]{pulseRev}
Wang, Z., Ma, Y., Cheng, F., Yang, L., 2010. Review of pulse-coupled neural
  networks. Image Vis. Comput. 28, 5--13.

\bibitem[{Werner(2010)}]{wernerFractais}
Werner, G., 2010. Fractals in the nervous system: conceptual implications for
  theoretical neuroscience. Front. Physiol. 1, 15.

\bibitem[{Yokoi et~al.(1985)Yokoi, de~Oliveira, and Salinas}]{yokoi1985}
Yokoi, C. S.~O., de~Oliveira, M.~J., Salinas, S.~R., 1985. Strange attractor in
  the ising model with competing interactions on the cayley tree. Phys. Rev.
  Lett. 54(3), 163--166.

\bibitem[{Zou et~al.(2009)Zou, Guan, and Lai}]{zouMapPulse}
Zou, H., Guan, S., Lai, C.-H., 2009. Dynamical formation of stable irregular
  transients in discontinuous map systems. Phys. Rev. E 80, 046214.

\end{thebibliography}




%% \bibitem must have one of the following forms:
%%   \bibitem[Jones et al.(1990)]{key}...
%%   \bibitem[Jones et al.(1990)Jones, Baker, and Williams]{key}...
%%   \bibitem[Jones et al., 1990]{key}...
%%   \bibitem[\protect\citeauthoryear{Jones, Baker, and Williams}{Jones
%%       et al.}{1990}]{key}...
%%   \bibitem[\protect\citeauthoryear{Jones et al.}{1990}]{key}...
%%   \bibitem[\protect\astroncite{Jones et al.}{1990}]{key}...
%%   \bibitem[\protect\citename{Jones et al., }1990]{key}...
%%   \harvarditem[Jones et al.]{Jones, Baker, and Williams}{1990}{key}...
%%


% \end{thebibliography}

\end{document}